\documentstyle[12pt,a4,bbbold,psfig,russian]{article}

\newcommand{\grad}{\mathop{\rm grad}\nolimits}
\def\dfrac#1#2{{\displaystyle#1\over\displaystyle#2}}
\def\slantfrac#1#2{\hbox{$\,^{#1}\!/_{#2}$}}

\def\apgt{\ {\raise-.5ex\hbox{$\buildrel>\over\sim$}}\ }
\def\aplt{\ {\raise-.5ex\hbox{$\buildrel<\over\sim$}}\ }
\newcommand{\Msyr}{$M_\odot~\mbox{yr}^{-1}$}
\def\Ms{M_{\odot}}
\def\Porb{P_{orb}}
\def\GWR{{\mbox{GWR}}}
\def\MSW{{\mbox{MSW}}}
\def\LOSS{{\mbox{LOSS}}}
\def\EXCH{{\mbox{EXCH}}}

\newcommand{\myv}{\vphantom{\frac11}}
\newcommand{\diver}{\mathop{\rm div}\nolimits}
\newcommand{\dd}{\mathop{\rm ~d}\nolimits}
\newcommand{\ddd}{\mathop{\rm \raisebox{0pt}{d}}\nolimits}

\begin{document}

\language=0

\title{Non-conservative Evolution\\ of Cataclysmic Variables}

\author{
A.V.Fedorova$^1$, D.V.Bisikalo$^1$, A.A.Boyarchuk$^1$,\\
O.A.Kuznetsov$^2$, A.V.Tutukov$^1$,
L.R.Yungelson$^1$\\[0.3cm]
$^1$ {\it Institute of Astronomy of the Russian Acad. of Sci.,
Moscow}\\
{\sf afed@inasan.rssi.ru; bisikalo@inasan.rssi.ru;
aboyar@inasan.rssi.ru;}\\
{\sf atutukov@inasan.rssi.ru; lry@inasan.rssi.ru}\\[0.3cm]
$^2$ {\it Keldysh Institute of Applied Mathematics, Moscow}\\
{\sf kuznecov@spp.keldysh.ru}\\[0.3cm]
}
\date{}
\maketitle

\begin{abstract}
{\large\bf Abstract}---We suggest a new mechanism to account for
the loss of angular momentum in binaries with non-conservative
mass exchange. It is shown that in some cases the loss of matter
can result in increase of the  orbital angular momentum of a
binary. If included into consideration in evolutionary
calculations, this mechanism appreciably extends the range of
mass ratios of components for which mass exchange in binaries is
stable. It becomes possible to explain the existence of some
observed cataclysmic binaries with high donor/accretor mass
ratio, which was prohibited in conservative evolution models.

\end{abstract}

\section{Introduction}

Cataclysmic variables (CVs) are close binary systems consisting
of a low-mass main-sequence (MS) star that fills its Roche lobe
and a white dwarf. Main-sequence star (mass-donating star,
usually referred to as star 2 or secondary) loses matter through
the vicinity of the inner Lagrangian point $L_1$. White dwarf
(accretor star, referred to as star 1 or primary) accretes at
least a fraction of this matter via accretion disk or via
accretion columns in the polar zones in the case when white
dwarf has a strong magnetic field.

Physics and evolution of CVs as well as evolution of similar to
them low-mass X-ray binaries were investigated starting from the
late sixties (see, e.g., [\ref{RJW82}--\ref{fed97}] and
references therein). These studies were motivated by the
recognition of the fact that evolution of CVs is determined by
losses of orbital angular momentum due to radiation of
gravitational waves [\ref{kmg62},\ref{p67}] and magnetically
coupled stellar wind [\ref{sch62}--\ref{vz81}]. A number of
studies investigated the influence of the loss of angular
momentum associated with the material outflow from the system on
the CV evolution (see, e.g., [\ref{yu73}--\ref{kk95}]). However,
in the absence of gas dynamical simulations of the mass transfer
in the binaries, these processes were considered under the
parametric approach.

Recent three-dimensional (3D) gas dynamical simulations of the
structure of gaseous flows in semi-detached binaries proved an
important role of a circumbinary envelope (see, e.g.
[\ref{bbkc98b},\ref{bbkc98c}]). These calculations also show
that during mass exchange a significant fraction of matter
leaves the system. The present work is mainly devoted to the
numerical investigation of the evolution of CVs using the data
of 3D gas dynamical calculations on the losses of mass and
angular momentum from  close binaries.  The main attention is
paid to the stability of mass exchange against runaway mass loss
and its dependence on the donor to accretor mass ratio
$q=M_2/M_1$.

It is known that the loss of matter by the donor star  results
in the violation of its hydrostatic and thermal equilibrium.
Hydrostatic equilibrium is restored adiabatically, i.e. in
dynamical time scale, while thermal equilibrium is restored in
thermal time scale. This transition to a new equilibrium state
leads to a change of donor radius. The sign of radius variation
depends on the convective and radiative stability of outer
envelope of the star. For stars with masses $M\aplt \Ms$ with
deep convective envelope as well as for white dwarfs mass loss
results in increase of the  radius, while for stars with
radiative envelopes it leads to the shrinkage of the star. Mass
exchange in close binary is unstable when in the course of
evolution mass-losing star tends to overfill the Roche lobe. It
can occur when radius of the donor $R_2$ increases faster (or
decreases slower) than the effective radius of Roche lobe
$R_{RL}$.\footnote{The effective radius of the Roche lobe
$R_{RL}$ is determined as the radius of a sphere with a volume
equal to the volume of the Roche lobe.} The case when the radius
of the donor increases while the radius of the Roche lobe
decreases is also possible. Thus, the question of stability of
mass exchange against runaway is determined by the balance of
derivatives $\partial R_2/\partial M_2$\ and $\partial
R_{RL}/\partial M_2$. It means that  unstable mass exchange is
possible even for stars which contract when losing matter.

The study of conditions of stable mass exchange is motivated by
two related problems: i) it is necessary to explain why $\sim$
10\% of CVs with well-determined masses of components have
combination of donor mass and mass ratio of components which is
``forbidden'' in the evolutionary models based on conventional
assumptions on the conservation of angular momentum of a binary;
ii) for population synthesis studies of CVs it is necessary to
distinguish progenitors of CVs among all binaries containing a
white dwarf and a low-mass companion (i.e., specify in which
binaries of this type stable mass exchange is possible). A
problem similar to the last one appears in the studies of
low-mass X-ray binaries.

The paper is organized as follows. In Section~2 we describe
major factors affecting the evolution of CVs and conditions of
stable mass exchange in binaries. The reasons which motivated us
to rule out the conservative approximation for mass exchange are
considered in Section~3. Sections~4 and~5 describe some results
of  3D gas dynamical simulations of the flow structure in
semi-detached binaries and introduce the model for description
of angular momentum losses due to non-conservative mass
exchange. Results of evolutionary calculations for conservative
and ``non-conservative'' models are compared in Section~6.
Criteria of stable mass exchange for ``non-conservative'' model
with mass and angular momentum loss are determined in Section~7.
Section~8 summarizes the main results of the work.

\section{Major factors affecting the evolution of CVs}

In the context of the present work the term ``stable mass
exchange'' means dynamical stability. In other words, in the
course of mass exchange the donor star can be out of thermal
equilibrium and mass transfer rate may exceed the rate
corresponding to the Kelvin time scale
($\stackrel{\bullet}{M}\,\simeq3\times10^{-7}RL/M$, where radius
$R$, luminosity $L$, and mass of the star $M$ are in solar units
and mass transfer rate $\stackrel{\bullet}{M}$ is in \Msyr). Let
us define unstable mass exchange as the situation when the
radius of the donor changes faster than the effective radius of
Roche lobe: $\stackrel{\bullet}
{R}_2\,>\,\stackrel{\bullet}{R}_{RL}$. Using the derivatives of
the radii w.r.t. mass of the donor  we can rewrite the condition
of the stability of mass transfer as:

\begin{equation}
\label{dr2r}
\zeta_\star\equiv\frac{\partial \ln R_2}{\partial \ln M_2}
> \frac{\partial \ln R_{RL}}{\partial \ln M_2}\equiv\zeta_{RL}\,.
\end{equation}
The effective radius of Roche lobe can be estimated, for
instance, using interpolation formula given by Eggleton
[\ref{eggl83}]:

$$
R_{RL} \approx 0.49 A \frac{q^{2/3}}{0.6q^{2/3}+\ln
(1+q^{1/3})}\,,
$$
where $A$ is the semimajor axis of the orbit. For  $q\aplt1$ the
approximation suggested by Paczy\'nski [\ref{pacyn71}] is more
convenient:

\begin{equation}
\label{rrp}
R_{RL} \approx \frac {2}{3^{4/3}}A
\left(\frac{q}{1+q}\right)^{1/3}\,.
\end{equation}
Derivative of the Roche lobe radius can be written as

\begin{equation}
\label{drra}
\frac{\partial \ln R_{RL}}{\partial \ln M_2}=
\frac{\partial \ln A}{\partial \ln M_2}
+\frac{\partial \ln (R_{RL}/A)}{\partial \ln M_2}\,.
\end{equation}
It is seen from (\ref{rrp}) and (\ref{drra}) that the variation
of effective radius of Roche lobe is defined by the change of
donor  and accretor masses $M_2$ and $M_1$ as well as by the
change in the separation $A$, and, hence, it is influenced by
the loss of mass and angular momentum from the system. As a
rule, only systemic angular momentum is considered as the
orbital momentum, i.e. the sum of angular momenta of two stars,
which are considered as point masses. Momenta of the spin of the
components, momentum of the accretion disc (if it's present), as
well as  momentum of gaseous flows inside the system as a rule
are not included into the total momentum.  Deviation of the
momentum of the donor from that of the point mass object (which
can be significant, see [\ref{kuz95}]) usually is not taken into
account as well. Such a simplified approach is compelled, since
it is very difficult to include all the abovementioned factors
into calculations. Having in mind all these simplifications, one
gets for the orbital momentum of a binary system with the
circular orbit

\begin{equation}
\label{j}
J=\sqrt{\dfrac{GM_1^2M_2^2A}{M_1+M_2}}\,,
\end{equation}
where $G$ is the gravitational constant.

According to the widely accepted models, the evolution of
cataclysmic binaries is determined by the loss of  angular
momentum from the system via gravitational waves radiation (GWR)
and/or magnetic stellar wind (MSW) of the donor as well as the
mass exchange between components. In the standard models of
evolution it's accepted that the mass exchange among components
does not change the systemic angular momentum, and influences it
indirectly -- through a possible loss of mass from the system
and angular momentum ablation by leaving matter (term with
subscript LOSS in Eq. \ref{dj} below).  Therefore, within
conventional models, the equation for the change of orbital
momentum can be written as:

\begin{equation}
\label{dj}
\frac{dJ}{dt} = \left(\frac{\partial J}{\partial
t}\right)_\GWR + \left(\frac{\partial J}{\partial
t}\right)_\MSW + \left(\frac{\partial J}{\partial
t}\right)_\LOSS\,.
\end{equation}
Let us consider the terms of expression (\ref{dj}):\\

\bigskip

{\it 1) Loss of the systemic angular momentum due to GWR}\\

The change of the systemic orbital momentum as a result of GWR
is given by a formula (see, e.g., [\ref{ll62}]):

\begin{equation}
\label{djgr}
\left(\frac{\partial \ln J}{\partial t}\right)_\GWR=
-\frac{32G^3}{5c^5}\frac{M_1M_2(M_1+M_2)}{A^4}\,,
\end{equation}
where $c$ is the speed of light. Intensity of GWR is
characterized by a very strong dependence on the separation of
components and respectively on the orbital period. This process
is essential for short-period systems with
$\Porb\aplt10^{\mbox{h}}$, since only then the time scale of
angular momentum loss becomes shorter than the Hubble time.\\

\bigskip

{\it 2) Loss of the systemic angular momentum due to MSW of the
donor}\\

A mechanism for the loss of the angular momentum of the system
by means of the MSW was suggested by Schatzman [\ref{sch62}] and
by Mestel [\ref{mes68}]. If the star has a convective envelope
and, subsequently, a surface magnetic field, its own axial
rotation is braked by the magnetic stellar wind, and the angular
momentum loss rate can be essential even for small mass loss
rate. The subsequent synchronization of the  spin of the donor
and  orbital rotation caused by tidal interaction between
components results in the loss of the orbital momentum of the
system and reduction of the separation of components $A$. To
take this effect into account, Verbunt and Zwaan [\ref{vz81}]
extrapolated the data on rotation of F-type stars
[\ref{skum72}] to the K and M type components of CVs and
suggested a widely used semi-empirical formula which allows to
determine the temporal behavior of the orbital momentum:

\begin{equation}
\label{djmsw}
\left(\frac{\partial\ln J}{\partial t}\right)_\MSW =
-0.5\times10^{-28}\cdot Ck^2G
\frac{(M_1+M_2)^2}{M_1}\frac{R_2^4}{A^5}\,,
\end{equation}
were $k$ is the radius of gyration of the donor star; $C$ is a
numerical factor determined from the comparison of theoretical
calculations and observational data. Well-defined reduction of
magnetic field for stars with masses less than $\sim0.3\Ms$
permits to suggest that vanishing of a radiative core of a star
when its mass decreases to this threshold leads to an abrupt
``switch-off'' of the so-called ``$\alpha\Omega$''-dynamo
mechanism that is responsible for the generation of stellar
magnetic fields [\ref{sr83}]. At the instant of cessation (or
sharp reduction) of MSW the time scale of angular momentum loss
[defined by Eq.~(\ref{djmsw})] is shorter than the thermal time
scale of a donor with $M\sim0.3\Ms$.   Thus, the latter is out
of thermal equilibrium and its radius is larger than the radius
of thermally equilibrium star of the same mass.\footnote{For
discussion on the response of star on mass loss in different
time scales see, e.g., [\ref{w85},\ref{pol96}].} When the action
of MSW and corresponding loss of angular momentum stop, the rate
of reduction of the semimajor axis declines, the donor shrinks
down to the equilibrium radius and loses the contact with Roche
lobe [\ref{sr83}]. Meanwhile, the system continues to lose
angular momentum due to GWR, components continue to approach
each other and after some time the optical star fills its Roche
lobe again. Further evolution of the system is determined by the
loss of angular momentum via GWR.

Cessation of MSW after the star becomes completely convective is
widely adopted hypothesis explaining satisfactorily the
so-called orbital period gap in CVs. The value of coefficient
$C$ in Eq.~(\ref{djmsw}) can be determined by comparison of
theoretical width of period gap and observational one. In the
present study we adopted $C=3.0$ in accordance to
[\ref{ft94}].\\

\bigskip

{\it 3) Loss of the systemic angular momentum due to mass
loss}\\

Mass loss from  the system is mainly considered as a parameter
fine tuning of which permits to get an agreement with
observational minimal period of CVs (e.g. [\ref{RJW82}]). More
definite assumptions are made only for specific cases when
examining evolution of the systems, in which: i) the rate of
accretion on white dwarf is limited by the hydrogen burning rate
($\sim10^{-7}\div10^{-6}$\,\Msyr), and the mass loss rate by the
donor is much higher than this limit, but does not exceed
Eddington limit for the dwarf (that is close to
$1.5\times10^{-5}$\,\Msyr~, see, e.g.,
[\ref{hkn96},\ref{yl98}]), the mass excess being lost by means
of stellar wind [\ref{kh94}]; ~ii) mass is lost due to outbursts
occurring on white dwarf (see, e.g., [\ref{yu73},
\ref{yuetal96}--\ref{skr98}])\footnote{The latter two  studies
considered also the loss of momentum due to interaction of the
donor with the envelope of a Nova.}. In both cases it is usually
assumed that the specific angular momentum of the matter leaving
the system is equal to the specific momentum of the accretor.

To describe the loss of mass and momentum out of the system, it
is convenient to introduce parameters describing the degree of
non-conservativity of the evolution respective to the mass
[\ref{ty71}]:

$$
\beta=-\frac{\stackrel{\bullet}{M}_1}{\stackrel{\bullet}{M}_2}
=-\frac{\partial M_1}{\partial M_2}
$$
and to the angular momentum:\footnote{\label{ftnt}Sometimes
specific angular momentum lost from the system is measured in
units of $\Omega A^2$\ and used instead of $\psi$:
$$
\alpha=\frac{\stackrel{\bullet}{J}}{\stackrel{\bullet}{M}}
\left/\Omega A^2\right.=
\frac{\stackrel{\bullet}{J}}{\stackrel{\bullet}{M}}
\left/\frac{J}{~\mu~}\right.=
\frac{q}{(1+q)^2}\psi\,,\qquad
\mu=\frac{M_1M_2}{M}\,.
$$
}

\begin{equation}
\label{eqpsi}
\psi=\frac{\stackrel{\bullet}{J}}{\stackrel{\bullet}{M}}
\left/\frac{J}{M}\right.
=\frac{\partial \ln J}{\partial \ln M}\,,
\end{equation}
where $M=M_1+M_2$.

With these parameters, the expression for the loss of momentum
by the matter leaving the system gets the following form:

$$
\left(\frac{\partial J}{\partial t}\right)_\LOSS=
(1-\beta)\stackrel{\bullet}{M}_2\psi\frac{J}{M}\,.
$$

In the calculations of the cataclysmic binary evolution an
equation for the variation of the semiaxis of the orbit  is used
directly, instead of the  equation for  the variation of
momentum over time.  Let us differentiate the expression
(\ref{j}), substitute the result into  equation (\ref{dj}), and
use the relation
$\stackrel{\bullet}{M}_1=-\beta\stackrel{\bullet}{M}_2$. Then
the equation for the variation of orbital semiaxis can be
written as:

\begin{equation}
\label{da}
\left(\frac{dA}{dt}\right) =
\left(\frac{\partial A}{\partial t}\right)_\EXCH+
\left(\frac{\partial A}{\partial t}\right)_\LOSS+
\left(\frac{\partial A}{\partial t}\right)_\GWR+
\left(\frac{\partial A}{\partial t}\right)_\MSW\,.
\end{equation}

\noindent
Here, index `EXCH' corresponds to the change of $A$ resulting
from the mass transfer between components. Note that while the
term $(\partial J/\partial t)_\EXCH$ is absent in the equation
(\ref{dj}), the appropriate term $(\partial A/\partial t)_\EXCH$
is present in the equation (\ref{da}), being a natural
consequence of the dependence of the orbital momentum both on
semimajor axis of the orbit and  masses of components. When mass
exchange is conservative, the evolution of the  orbital
separation is determined by the assumption that the mass
transfer does not change the orbital momentum of the system. In
the course of mass exchange the matter that initially had
specific angular momentum of the donor is transferred onto
accretor and finally gets the specific momentum of the accretor.
The assumption of conservation of the orbital momentum implies
that if the mass of the donor is lower than the  mass of
accretor, the excess momentum of accreted gas should transform
into the orbital one, and the mass exchange acts in the
direction of increasing the orbital semiaxis. If the donor is
more massive than the accretor, the lacking momentum of the
accreted gas should be taken from the orbital momentum,
therefore, the mass exchange acts in the direction of shrinking
of the binary system.

The numerical investigation of the evolution of cataclysmic
binaries consists of simultaneous calculations of the evolution
of the donor  and of the variation of the orbital separation in
time. Let us consider the processes that determine the evolution
of the orbital semiaxis. Using parameters $\beta$ and $\psi$, we
can write formulae for the terms of equation (\ref{da}) (here
masses and distances are given in solar units and time is given
in years):

\begin{equation}
\label{daex}
\left(\frac{\partial A}{\partial t}\right)_\EXCH=
2A\frac{M_2-M_1}{M_1 M_2}\beta\stackrel{\bullet}{M}_2\,,
\end{equation}

\begin{equation}
\label{daloss}
\left(\frac{\partial A}{\partial t}\right)_\LOSS=
-(1 - \beta) A
\frac{2M_1(1-q\psi)+M_2}{M_2(M_1+M_2)}\stackrel{\bullet}{M}_2\,,
\end{equation}

\begin{equation}
\label{dagr}
\left(\frac{\partial A}{\partial t}\right)_\GWR=
-1.65\times10^{-9}\frac{M_1M_2(M_1+M_2)}{A^3}\,,
\end{equation}

\begin{equation}
\label{damsw}
\left(\frac{\partial A}{\partial t}\right)_\MSW =
-6\times10^{-7}C\frac{(M_1+M_2)^2}{M_1}
\left(\frac{R_2}{A}\right)^4\,.
\end{equation}

It is seen from Eqs~(\ref{daex})--(\ref{damsw}) that changing of
$R_{RL}$ is mainly determined by the value of $q$. On the other
hand, changing of $R_2$ depends on $M_2$ and
$\stackrel{\bullet}{M}_2$. Thus the most convenient way for
studying of limits of stable mass exchange is analysis of ``ratio
of donor-star mass to accretor mass $q$ -- donor-star mass
$M_2$'' diagram.

In a number of cases one may simplify expression (\ref{drra})
and obtain analytical formulae for the derivative of the Roche
lobe radius. In particular, when the angular momentum changes
due to GWR and MSW can be neglected, one may deduce from
(\ref{daex}) and (\ref{daloss}) the expression for the change of
separation $A$ [the first term in the formula (\ref{drra})] as a
function of mass ratio $q$ and the parameters of the loss of
matter and momentum $\beta$ and $\psi$:

\begin{equation}
\label{dabpsi}
\frac{\partial\ln A}{\partial\ln M_2}=\frac{
2\psi(1-\beta)q+\beta q+2\beta q^2-q-2}{1+q}\,.
\end{equation}
Using the expression (\ref{rrp}) for $R_{RL}$ we can express the
second term of (\ref{drra}) as:

\begin{equation}
\label{dabpsi1}
\frac{\partial \ln (R_{RL}/A)}{\partial \ln M_2}
=\slantfrac{1}{3}-\slantfrac{1}{3}\frac{(1-\beta)q}{1+q}\,.
\end{equation}
Using (\ref{dabpsi}) and (\ref{dabpsi1}), for the case of
totally conservative mass exchange ($\beta=1$), we obtain:

\begin{equation}
\label{case1}
\frac{\partial\ln R_{RL}}{\partial \ln M_2} =
2q-\slantfrac{5}{3}\,.
\end{equation}

From the same equations, for the so-called `Jeans mode' of mass
loss, when the mass exchange in the system does not occur, but
the donor loses its mass due to stellar wind and matter leaves
the system carrying away  specific angular momentum of the donor
($\beta=0$, $\psi=q$), we obtain:

\begin{equation}
\label{case2}
\frac{\partial \ln R_{RL}}{\partial \ln M_2} =
\slantfrac{1}{3}
\frac{1-3q}{1+q}\,.
\end{equation}

For the case of stellar wind from the accretor when all matter
lost by the donor is transferred onto accretor but then leaves
the system carrying away specific angular moment of the accretor
(formally, one should take here $\beta=0$ and $\psi=q^{-1}$) we
obtain:

\begin{equation}
\label{case3}
\frac{\partial \ln R_{RL}}{\partial \ln M_2} =
\frac{2q^2-q-\slantfrac{5}{3}}{1+q}\,.
\end{equation}

The comparison of the rates of variation of the radius of the
donor and effective radius of the Roche lobe  (\ref{drra})
determines the state of stability/instability of mass exchange
in any phase of evolution. It is possible to determine the
boundaries of stable mass exchange region by two different
methods:

\begin{itemize}

\item It is possible to calculate derivatives of the radius of
the donor w.r.t. its mass directly for different
$\stackrel{\bullet}{M}_2$ and to compare it with derivatives of
the effective radius of Roche lobe depending on $M_1$, $M_2$,
and $A$. This method was used, for instance, in [\ref{tfy82}]
and it permits to restrict a region of $q-M_2$ diagram where
mass exchange is dynamically stable and also to detect the
regions where mass exchange should occur in different time
scales: thermal, nuclear, or in the time scale of angular
momentum loss.

\item Alternatively, a requirement that the rate of accretion
should not exceed the Eddington limit for the dwarf can be
accepted as a criterion of stable mass exchange. This assumption
seems to be more reasonable since evolutionary calculations
often suggest very high rate of mass loss during very initial
moments of time after overfilling the Roche lobe, while after
that $\stackrel{\bullet}{M}$ falls off rapidly. In fact, this
approach involves a restriction on dynamical stability of mass
transfer but permits loss of matter in thermal time scale.

\end{itemize}

\section{Stable mass exchange region in conservative on mass
model of the CV evolution}

Figure~\ref{ris1} depicts the boundary of the stable mass
exchange region in conventional conservative model of the
evolution of CV. Boundary 1 was found in [\ref{tfy82}], and
boundary 2 was found in [\ref{kool92}]. These boundaries
coincide well for of donors of low mass but differ a little for
large ones.  Unfortunately, in [\ref{kool92}] the procedure of
the calculation of the boundary isn't described  in detail. We
may assume that this discrepancy is due to the differences in
the codes, in a different treatment of possible weakening of
donor MSW at masses greater than $\sim 1 \Ms$, and different
techniques for computation of stellar radii derivatives in this
region.

\renewcommand{\thefigure}{1}
\begin{figure}[p]
\centerline{\hbox{\psfig{figure=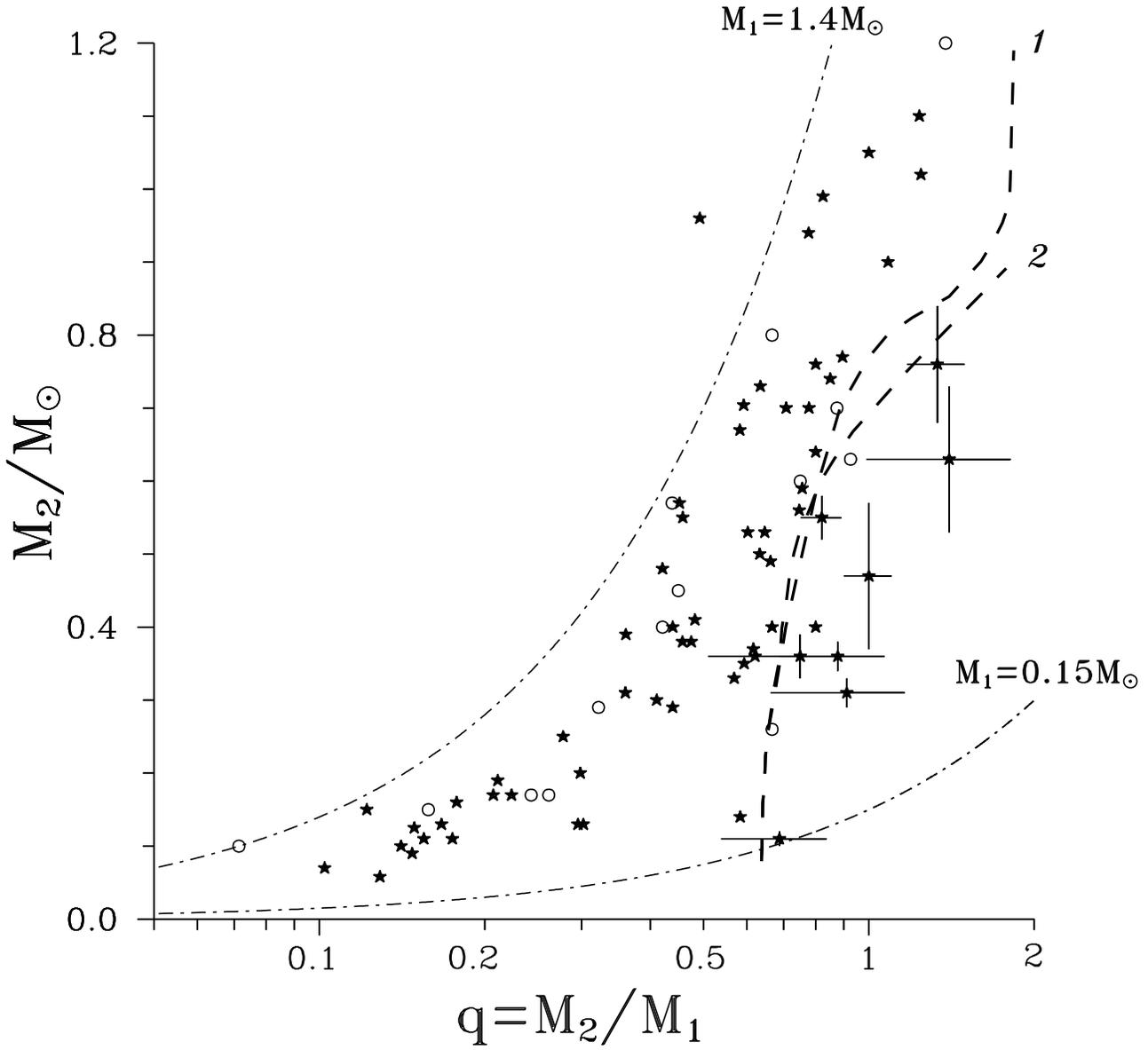,width=17cm}}}
\caption{\small ``Mass ratio of donor to accretor  $q$
-- donor mass $M_2$'' diagram. Asterisks $\star$ and circles
$\circ$ show observed stars from the catalogue of Ritter \& Kolb
[\ref{cat97}] (marker $\circ$ indicates that data are not
reliable). For some CVs error bars are also shown. Dashed lines
-- boundary of stable mass exchange region in the conservative
model according to [\ref{tfy82}] (dashed line 1) and according
to [\ref{kool92}] (dashed line 2). Dot-dashed lines show the
upper ($M_1\leq1.4\Ms$) and lower ($M_1\geq0.15\Ms$) cutoffs of
white dwarf mass.}
\label{ris1}
\end{figure}

\renewcommand{\thefigure}{1a}
\begin{figure}
\centerline{\hbox{\psfig{figure=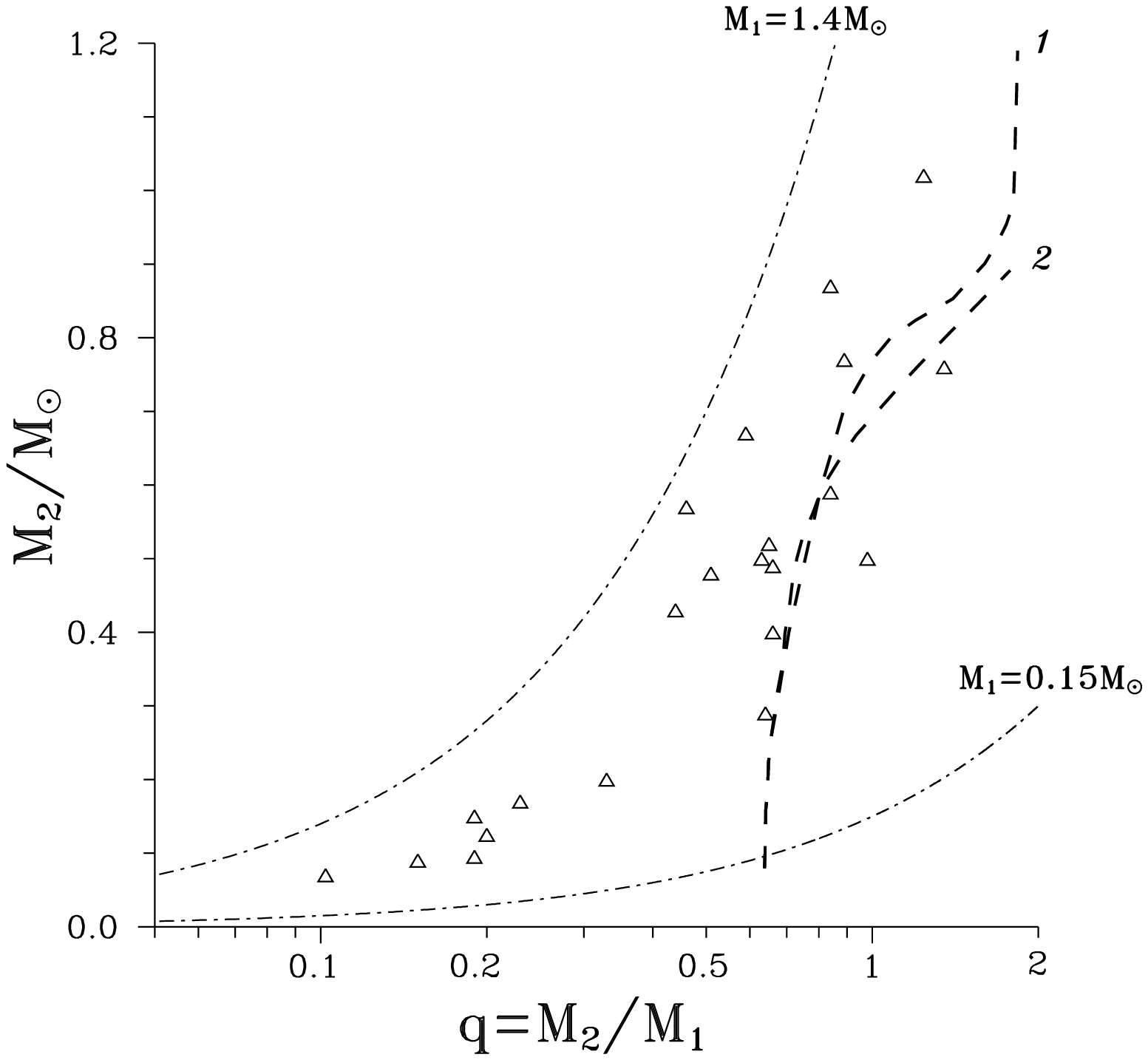,width=17cm}}}
\caption{\small The same as in Fig.~\ref{ris1} but with observed
stars from [\ref{smith_dhillon}].}
\end{figure}

Figure~\ref{ris1} also shows CVs with known masses of components
from the catalog [\ref{cat97}]. The total number of binaries in
the diagram is 80, and 11 of them are in the region of unstable
mass exchange. For 8 of these 11 systems uncertainties of
parameters are known, they are shown in  Fig.~\ref{ris1} as
error bars. It is seen that the location of stars in a region of
unstable mass exchange can not be explained by uncertainties in
masses of components. In Fig.~\ref{ris1}a we plot CVs with the
best determined masses [\ref{smith_dhillon}], parameters for a
number of systems being different compared to the data from
[\ref{cat97}].  Nevertheless, even for well-determined systems
the problem of ``improper'' location of CVs remains -- 3 out of
22 systems lie in the ``forbidden'' region.

To explain this disagreement we suggest a model of CVs evolution
which involves loss of mass and angular momentum from the
system. This model is based on 3D gas dynamical simulations of
the flow structure in close binary systems.

\section {Calculations of the rate of an angular moment loss in
3D gas dynamical models}

The above-stated analysis proves that the rate of change of the
distance between components $A$ in evolutionary calculations
depends heavily on two parameters: the degree of
non-conservativity of mass exchange w.r.t. the matter $\beta$
and w.r.t. to the angular momentum $\psi$. These parameters can
be determined only by means of 3D gas dynamical simulations of
mass transfer in binary systems. These investigations were made
in [\ref{bbkc98b}] for a binary system with  mass ratio of
components $q=\slantfrac{1}{5}$, and in [\ref{bbkc98c}] for a
binary with $q=1$. Besides that, we have specially made a
numerical simulation of a binary system with $q=5$ to study the
influence of parameter $q$ on the flow structure.

Analysis of the obtained results proves that in the binary
systems considered by us for different values of parameter $q$
mass transfer is non-conservative and the non conservativity
degree w.r.t. the matter is $\beta\sim0.4\div0.6$. Simulations
of flows in the 3D numerical models prove also that the matter
leaves the system with a significant angular momentum. The non
conservativity degree of mass exchange w.r.t. the matter can be
easily determined directly from gas dynamical model, while the
determination of non conservativity degree w.r.t. the angular
momentum is a sophisticated problem. Let us consider the
equations which determines the transport of angular momentum in
the system. We should consider Navier-Stokes equations for a
correct analysis because viscosity plays a significant role in
the redistribution of angular momentum.  From the steady-state
gas dynamical equations in a rotating frame we can deduce the
equation for the angular momentum transfer

$$
u\frac{\partial \lambda}{\partial x}
+v\frac{\partial \lambda}{\partial y}
+w\frac{\partial \lambda}{\partial z}
+\frac{1}{\rho}\left(\myv({\bmath r}-{\bmath r}_{CM})\times
\diver {\cal P}\right)_z=-\left(\myv({\bmath r}-{\bmath
r}_{CM})\times\grad \Phi\right)_z\,,
$$
where, as usually, ${\bmath v}=(u,v,w)$ is the velocity vector,
$\rho$ is the density, $\Phi$ is the Roche potential, $\Omega$
is the orbital angular velocity, ${\cal P}$ is the stress
tensor, ${\bmath r}_{CM}$ is the radius-vector of the center of
masses of the system, and angular momentum $\lambda$ (in a
laboratory, i.e. inertial or observer's frame) is defined by
expression:

$$
\begin{array}{l}
\lambda=(x-x_{CM})v-yu+\Omega\left(\myv(x-x_{CM})^2+y^2\right)\\
~\\
\qquad\qquad\qquad
=(x-x_{CM})\left(\myv v+\Omega (x-x_{CM})\right)-y(u-\Omega y)\,.
\end{array}
$$
Let us write down the equation for angular momentum transfer in a
divergent form:

$$
\diver(\rho\lambda{\bmath u})
+\left(\myv({\bmath r}-{\bmath r}_{CM})\times \diver {\cal
P}\right)_z= -\rho\left(\myv({\bmath r}-{\bmath
r}_{CM})\times\grad \Phi\right)_z\,.
$$
From this equation, we can obtain the integral law of change of
angular momentum as follows:

$$
\int\limits_\Sigma\rho\lambda{\bmath u}\cdot\ddd{\bmath n}
+\int\limits_\Sigma \left(\myv({\bmath r}-{\bmath r}_{CM})\times
{\cal P}\ddd{\bmath n}\right)_z=
\int\limits_V\rho\left(\myv({\bmath r}-{\bmath
r}_{CM})\times\grad\Phi\right)_z\dd V\equiv\Pi\,.
$$
The term $\Pi$ determines the change of the  angular momentum
due to non central nature of the force field, defined by the
Roche potential. By definition, the value

\begin{equation}
{\bmath F}_\lambda=\rho\lambda{\bmath u}+{\cal P}{\bmath
{r'}}\,,
\qquad\mbox{where}\qquad {\bmath {r'}}=(-y,x-x_{CM},0)
\label{potok}
\end{equation}
is the specific flux of angular momentum $\lambda$. As a result,
for the steady-state case we obtain:

$$
-\int\limits_{\Sigma_1}{\bmath F}_\lambda\cdot\ddd{\bmath n}=
\int\limits_{\Sigma_2}{\bmath F}_\lambda\cdot\ddd{\bmath n}
+\int\limits_{\Sigma_3}{\bmath F}_\lambda\cdot\ddd{\bmath
n}-\Pi\,,
$$
where $\Sigma_1$ and $\Sigma_2$ are the boundaries of the donor
and the accretor, respectively, and $\Sigma_3$ is the outer
boundary of the calculation domain. By analogy to the flux of
matter

$$
\stackrel{\bullet}{M}
=-\int\limits_{\Sigma_3}{\bmath F}_m\cdot\ddd{\bmath n}
=-\int\limits_{\Sigma_3}\rho{\bmath u}\cdot\ddd{\bmath n}\,,
$$
one may estimate the flux of angular momentum from the system as

\begin{equation}
\stackrel{\bullet}{J}
=-\int\limits_{\Sigma_3}{\bmath F}_\lambda\cdot\ddd{\bmath n}\,,
\label{error}
\end{equation}
and to use it in the expression (\ref{eqpsi}), determining the
non conservativity parameter w.r.t. the angular momentum.

Note that numerical simulations were conducted for Eulerian
equations for in-viscid gas and not for Navier-Stokes equations.
Respectively, to calculate integral (\ref{error}) we use
isotropic term describing gas dynamical pressure in the
expression for the specific flux of angular momentum: ${\cal
P}_{\alpha\beta}=P\cdot\delta_{\alpha\beta}$.\footnote{Here, as
usual, $\delta_{\alpha\beta}$ is Kronecker's symbol:
$$
\delta_{\alpha\beta}=\left\{
\begin{array}{l}
1\,,\qquad\alpha=\beta\,,\\
0\,,\qquad\alpha\ne\beta\,.
\end{array}\right.
$$}
This substitution is correct because on the outer border where
the integral in (\ref{error}) is estimated, viscosity does not
play a significant role.

Using expression (\ref{error}) in the gas dynamical calculations
we obtained $\psi\sim6$ corresponding to $\alpha=0.83$ (see
definition of $\alpha$ in the footnote on page \pageref{ftnt})
for the binary system with $q=\slantfrac{1}{5}$ [\ref{bbkc98b}],
and $\psi\sim5$ that corresponds to $\alpha=1.25$
[\ref{bbkc98c}] for a binary  with $q=1.$ The last value is in
good agreement with the results of two-dimensional calculations
of binary system with equal masses of components
[\ref{sawada84}] where $\alpha=1.65$ was obtained.

However, the use of these estimates for $\psi$ in evolutionary
calculations proves that this rate of an angular moment loss is
so great that in the majority of binary systems the mass
exchange rapidly becomes a runaway process and the rate of mass
loss by the donor increases without any limit. Apparently, the
application of formula (\ref{error}) to the estimation of   the
angular momentum loss in the evolutionary calculations is not
quite correct. since conjunction of gas dynamical and
evolutionary models is not completely consistent. The considered
gas dynamical model does not take into account the change in the
position of stars in time and, hence, the change in the angular
momentum of the gas due to the non-central nature of the force
field is not compensated by respective change in the orbital
angular momentum.

In a more general gas dynamical model, in which the change in
the position of stars is taken into account, one may correctly
estimate the change of the systemic angular momentum in the form
of angular momentum flux through the external boundary for the
``donor + accretor + gas'' system. However, to take into account
such gas dynamical calculations appears rather complicated.
Such  estimates are possible only for a specific stage in the
life of a binary system. Moreover, thus obtained gas dynamical
results cannot be used directly in the standard evolutionary
models that ignore the presence of circumbinary envelope in the
system. Therefore, a simplified model for estimation of the
angular momentum loss developed on the basis of gas dynamical
calculations will be a promising approach in the nearest future.

\section{A simplified model for estimation of angular momentum
loss rate in semi-detached binaries}

Gas dynamical simulations of mass transfer in semi-detached
binaries confirm that outflow of matter from the donor occurs
through a small vicinity of the inner Lagrangian point $L_1$.
In this case specific angular momentum of lost matter (in
corotating frame) can be estimated as
$\lambda_{L_1}=\Omega\Delta^2$, where $\Delta=|x_c-x_{L_1}|$.
Respectively, the flow of angular momentum from the donor is
equal to

$$
F_\lambda^{(2)}=\lambda_{L_1}\stackrel{\bullet}{M}_2
=\Omega\Delta^2\stackrel{\bullet}{M}_2\,.
$$

Due to non-conservativity of mass exchange, only a fraction
$\beta$ of the matter outflowing through $L_1$ is accreted.
Specific angular momentum of accreted matter is equal to the
angular momentum of accretor (here we neglect the finite radius
of accretor and/or residual angular momentum of accreted matter,
and this approximation appears to be reasonable for CVs). It
means that accreted matter has zero angular momentum in the
rotating frame, and, hence, does not accelerate the rotation of
accretor. Thus the question on the efficiency of the angular
momentum transfer from the spin rotation of accretor to the
orbital rotation of the binary can be ignored. To summarize, we
can describe the  flow of accreted angular momentum as

$$
F_\lambda^{(1)}=\lambda_{accr}\beta\stackrel{\bullet}{M}_2
=\Omega\left(\frac{M_2}{M}\right)^2A^2\beta\stackrel{\bullet}{M}_2\,.
$$

These expressions for the angular momentum flows do not
contradict the general formula for specific flow of angular
momentum [see (\ref{potok})]. Indeed, viscosity does not play a
significant role neither in the vicinity of the surface of the
donor nor in the vicinity of the accretor where the matter has
already  lost its angular momentum and the flow has radial
direction in the rotating frame. It means, that in these cases
the stress tensor ${\cal P}$ is reduced to the isotropic
pressure as well, and does not deposit into the integral of
specific flow of angular momentum. Our assumptions permit to
write the expression for the flow of angular momentum lost from
the system in the following form:

$$
F_\lambda^{loss}=\eta\left(F_\lambda^{(2)}-F_\lambda^{(1)}\right)=
\eta\left(\Omega\Delta^2\stackrel{\bullet}{M}_2
-\Omega\left(\frac{M_2}{M}\right)^2
A^2\beta\stackrel{\bullet}{M}_2\right)\,,
$$
where $\eta$ is a parameter defining the fraction of the
circumbinary envelope angular momentum that is lost with the
matter leaving the system. Then $1\!-\!\eta$ is the fraction of
the circumbinary envelope angular momentum that returns back
into the system via tidal interaction. Henceforth we adopt for
the parameter $\eta$  a value of 1 that corresponds to the case
when the angular momentum of the circumbinary envelope is
totally lost with the matter leaving the system. In this case
the  specific angular momentum of the matter leaving the system
is equal to

$$
\lambda_{loss}=F_\lambda^{loss}/\stackrel{\bullet}{M}_{loss}
=\frac{\Omega\Delta^2\stackrel{\bullet}{M}_2
-\beta\left(\dfrac{M_2}{M}\right)^2 \Omega
A^2\stackrel{\bullet}{M}_2}
{(1-\beta)\stackrel{\bullet}{M}_2}
$$
or in the units of systemic specific angular momentum
($\lambda_{syst}=\Omega A^2M_1M_2/M^2$):

\begin{equation}
\label{XXXX}
\psi=\lambda_{loss}/\lambda_{syst}=
\frac{\left(f(q)-\dfrac{1}{1+q}\right)^2
-\beta\left(\dfrac{q}{1+q}\right)^2}{1-\beta}\cdot
\frac{(1+q)^2}{q}\,,
\end{equation}
where $f(q)=x_{L_1}/A$ is dimensionless distance from the center
of the donor to $L_1$.

We would like to stress that formula (\ref{XXXX}) was derived
under the assumption of aligned synchronous rotation of
components of the binary system. It is known that the loss of
angular momentum due to outflow of the matter from the vicinity
of $L_1$ can result in the development of a misalignment of the
vector of spin rotation of the donor and the vector of the
orbital rotation of the binary [\ref{matese83}]. The flow
structure in the binaries with misaligned asynchronous rotation
was studied by the authors in [\ref{bbkc98d}]. Having in mind
the simplified character of evolutionary models we don't
consider the effect of the misalignment in the present work.

\renewcommand{\thefigure}{2}
\begin{figure}[p]
\centerline{\hbox{\psfig{figure=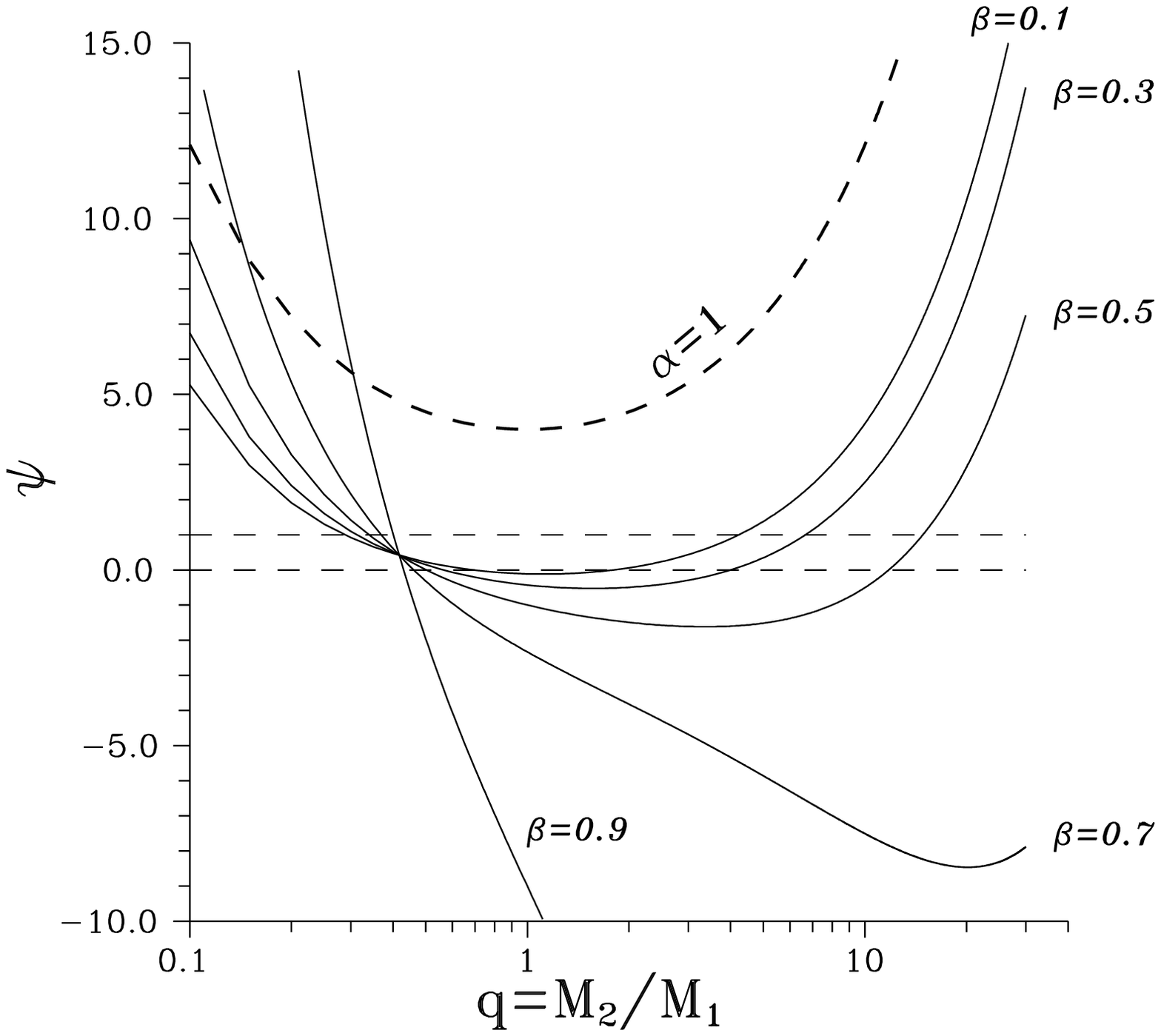,width=16cm}}}
\caption{\small Non-conservativity degree of the evolution
respective to angular momentum $\psi$ as a function of mass
ratio $q$ for various values of non-conservativity degree of the
evolution respective to mass $\beta$ [see Eq.~(\ref{XXXX})].
Thin dashed lines correspond to the values of $\psi=0$ and
$\psi=1$. Bold dashed line shows a dependence $\psi=(1+q)^2/q$
corresponding to the value $\alpha=1$ (see definition of
$\alpha$ in the footnote on p.  \pageref{ftnt}).}
\label{ris2}
\end{figure}

The graph of $\psi(q)$ is shown in Fig.~\ref{ris2} for various
values of non-conservativity degree w.r.t. the matter $\beta$.
The values $\psi=0$ and $\psi=1$ are also shown in the Figure by
dashed lines. In the region $\psi>1$ each unit of the matter
leaving the system carries away an angular moment that is higher
than the mean specific angular momentum of the system. This
results in diminishing of systemic specific angular momentum:

$$
\dfrac{d}{dt}\left(\dfrac{J}{M}\right)<0\,.
$$

In the region $0<\psi<1$ each unit of the matter leaving the
system carries away an angular moment that is lower than the
mean specific angular momentum of the system, and, hence,

$$
\dfrac{d}{dt}\left(\dfrac{J}{M}\right)>0\,.
$$

Figure~\ref{ris2} shows that the value of $\beta$ becomes
negative in a certain range of $q$ for the majority of cases of
non-conservative mass exchange (except $\beta\sim0$). This means
that the escape of the matter to infinity  causes redistribution
of the angular momentum in the ``donor + accretor + gas''
system. This results in the growth  of the total angular
momentum of the binary:  $\stackrel{\bullet}{J}\,>0$. It is
evident that such a ``pumping'' of angular momentum into the
system can stabilize the mass exchange in CV. Indeed,
consideration of orbit semiaxis change due to conservative mass
exchange shows that the binary system widens for $q<1$ only. For
the ``non-conservative'' model the range of $q$ corresponding to
the widening of binary due to mass and angular momentum losses
is larger -- for $\beta=0.5$ it happens for $0<q<2.8$.

Note that $\psi(q,\beta)$ dependence has a curios feature [see
Fig.~\ref{ris2} and formula (\ref{XXXX})] -- all curves
$\psi(q)$ for different values of $\beta$ pass through the same
point.  This point corresponds to the case when the center of
mass of the binary system is located exactly equidistantly
between $L_1$ and accretor, and in this case $\psi$ does not
depend on $\beta$.  Fortunately, this specific feature does not
influence  the analysis of solution.

\renewcommand{\thefigure}{3}
\begin{figure}[p]
\centerline{\hbox{\psfig{figure=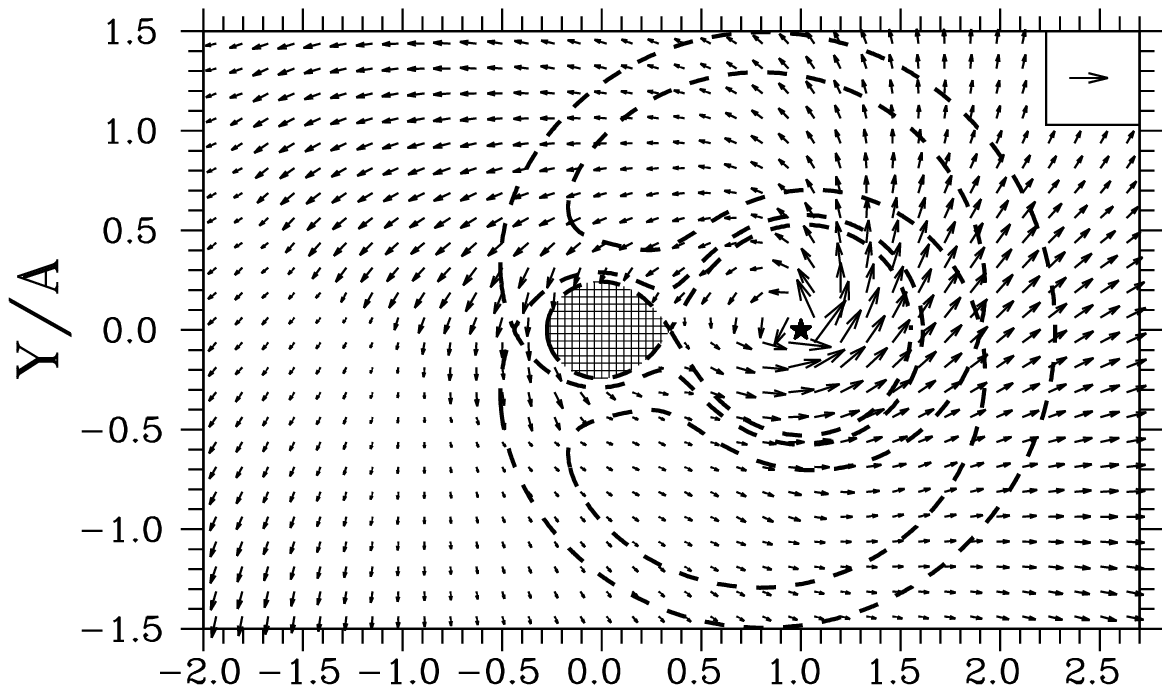,width=15cm}}}
\vspace{5mm}
\centerline{\hbox{\psfig{figure=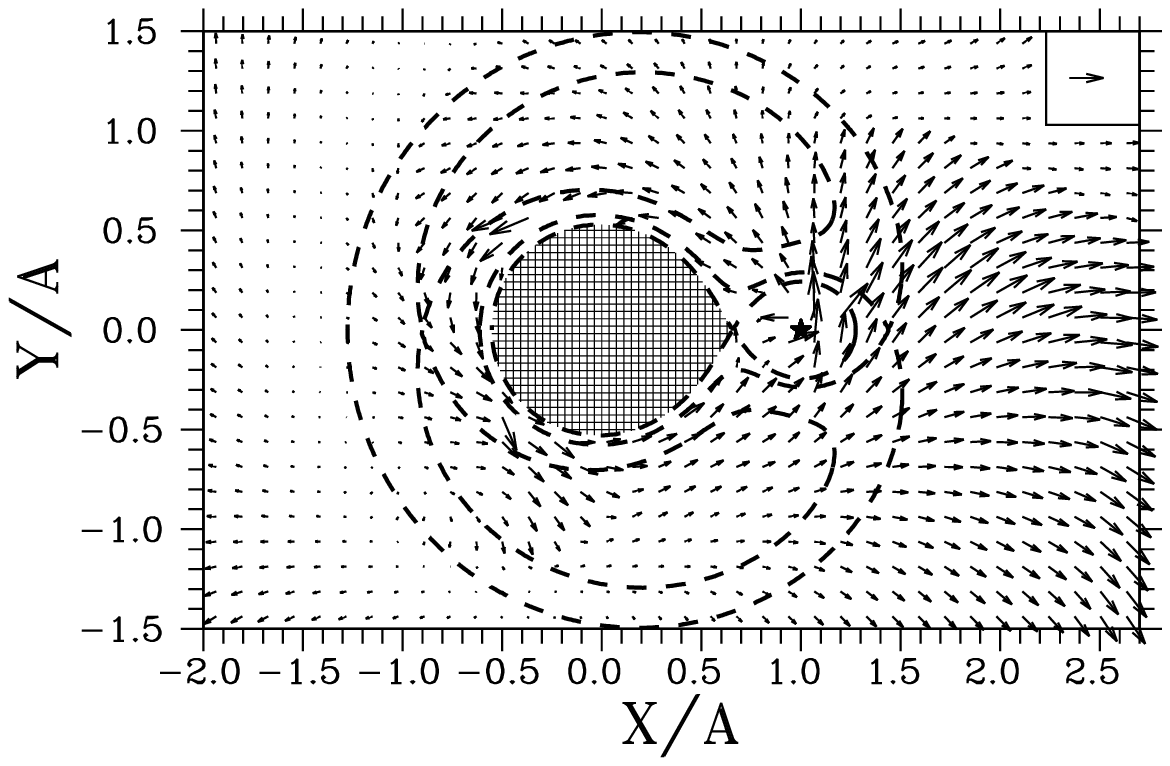,width=15cm}}}
\caption{\small Velocity vectors in equatorial plane for 3D gas
dynamical simulation of a binary system with $M_2/M_1=1:5$
($q=\slantfrac{1}{5}$, top panel) and with $M_2/M_1=5:1$ ($q=5$,
lower panel). Position of the accretor is marked by an asterisk
$\star$.  Shaded region marks the  donor. Dashed lines show
Roche equipotentials. A vector in the top right corner
corresponds to the velocity equal to $3A\Omega$.}
\label{ris3}
\end{figure}

Three-dimensional gas dynamical simulation of mass transfer in
CVs confirms conclusion that the outflow of matter from the
system can lead both to decrease and to increase of the total
angular momentum of a binary. The typical flow structures [see
Fig.~\ref{ris3}a,b] for binaries with $M_2/M_1=1:5$ and
$M_2/M_1=5:1$ demonstrate that in an inertial frame gas can have
different directions of rotation (in both figures orbital
rotation corresponds to the counter-clockwise direction).
Respectively, the change of angular momentum has opposite signs
in the obtained solutions -- for $q=\slantfrac{1}{5}$ the
systemic angular momentum  decreased, and for $q=5$ it
increased.

It seems reasonable to explain the different types of gas
dynamical solutions using the following qualitative speculation:

\begin{itemize}

\item The flow of the gas leaving the system consists of the
matter of the circumbinary envelope that has an angular momentum
large enough to overcome the gravitational  attraction of the
accretor. Initial velocity of this matter (in rotational frame)
has the same direction as the orbital movement of the accretor.

\item Velocity of the gas leaving the system  changes under the
action of the gravitational attraction of both components of the
binary system, centrifugal force, Coriolis force, and pressure
gradient. The change of azimuthal velocity is determined mainly
by the Coriolis force and gravitation forces, since centrifugal
force is axially-symmetrical and deviation of pressure gradient
from the axial symmetry is small as well. Action of Coriolis
force results in deflecting of the stream in the direction
opposite to that of the orbital movement. Consequently, this
permits to subdivide the whole flow leaving the system through
the vicinity of $L_2$ into two parts: the first part retains the
direction of original movement, and the direction of movement of
the second one is changed.

\item It is evident that in an axially-symmetrical gravitational
field the action of Coriolis force can not produce a solution
when the direction of gas movement in the inertial frame is
opposite to the orbital one. The situation is changed
drastically for the case of non-axially-symmetrical
gravitational field of a binary system. In this case
gravitational  attraction of the donor can accelerate the gas of
the stream and result in the solution with opposite direction of
movement in the inertial frame.

\item It follows from Eq.~(\ref{XXXX}) and Fig.~\ref{ris3} that
there is a range of values of $q$ where gas moves in the
direction opposite to the orbital one in the inertial frame.
These values of $q$ correspond to the cases when attraction of
the donor is large enough to accelerate the gas properly (left
end of the range), but not too large to produce accretion of gas
on the surface of the donor (right end of the range). The values
of $q$ out of this range correspond to the case when the flow
directed along the orbital movement prevails.

\end{itemize}

\noindent
Appearance of an additional range of values of $q$ where the
``non-conservative'' mass exchange is accompanied by  increase
of systemic angular momentum  results in expansion of the stable
mass exchange region. To illustrate this fact let us consider a
graph that is often used in the qualitative studies of the
stability of mass transfer in CV. This is a $q-\zeta$ graph with
curves corresponding to the analytical expressions for
derivatives of effective radius of the Roche lobe w.r.t. the
mass of the donor $\zeta_{RL}$, and straight lines corresponding
to the values of derivatives of the radius of the star w.r.t. to
its mass $\zeta_\star$, the latter being known from the
investigations of the internal structure of star (see, e.g.,
[\ref{tfy82}]). Mass exchange is stable for the range of $q$
where curve $\zeta_{RL}$ passes below the straight line
$\zeta_\star$ for the donor of corresponding type. This graph is
presented in Fig.~\ref{ris4}.  It shows the derivative of
effective radius of Roche lobe $\zeta_{RL}=\partial\ln
R_{RL}/\partial\ln M_2$ as a function of $q$ for a scenario in
which the loss of angular momentum is defined by formula
(\ref{XXXX}) for various values of the parameter $\beta$.
Dependencies $\zeta_{RL}(q)$ for the  scenario of the completely
conservative mass exchange and for the case when the outflowing
matter carries away specific angular momentum of the donor or
specific angular momentum of accretor are shown as well. The
same Fig.~\ref{ris4} also shows the values of derivative
$\zeta_\star=\partial\ln R_2/\partial\ln M_2$ for completely
convective and degenerate stars
($\zeta_\star=-\slantfrac{1}{3})$, for thermally equilibrium
stars of MS with $M\sim \Ms$ ($\zeta_\star=0.6)$, and for
subgiants with degenerate low-mass helium cores
($\zeta_\star=0)$. The analysis of Fig.~\ref{ris4} proves that
the outflow of matter and loss of angular momentum from the
system in accordance to law (\ref{XXXX}) stabilize mass exchange
in a wider range of $q$  compared to the conventional models of
conservative mass exchange. It is also worth noting that for
conservative scenario of mass exchange its time scale for stars
with masses $\sim \Ms$ is determined by the time scale of the
loss of angular momentum or the nuclear time scale for $q\aplt
1.2$ [\ref{tfy82}], while for above-considered
``non-conservative'' model this limit on $q$ is sufficiently
higher.

\renewcommand{\thefigure}{4}
\begin{figure}[p]
\centerline{\hbox{\psfig{figure=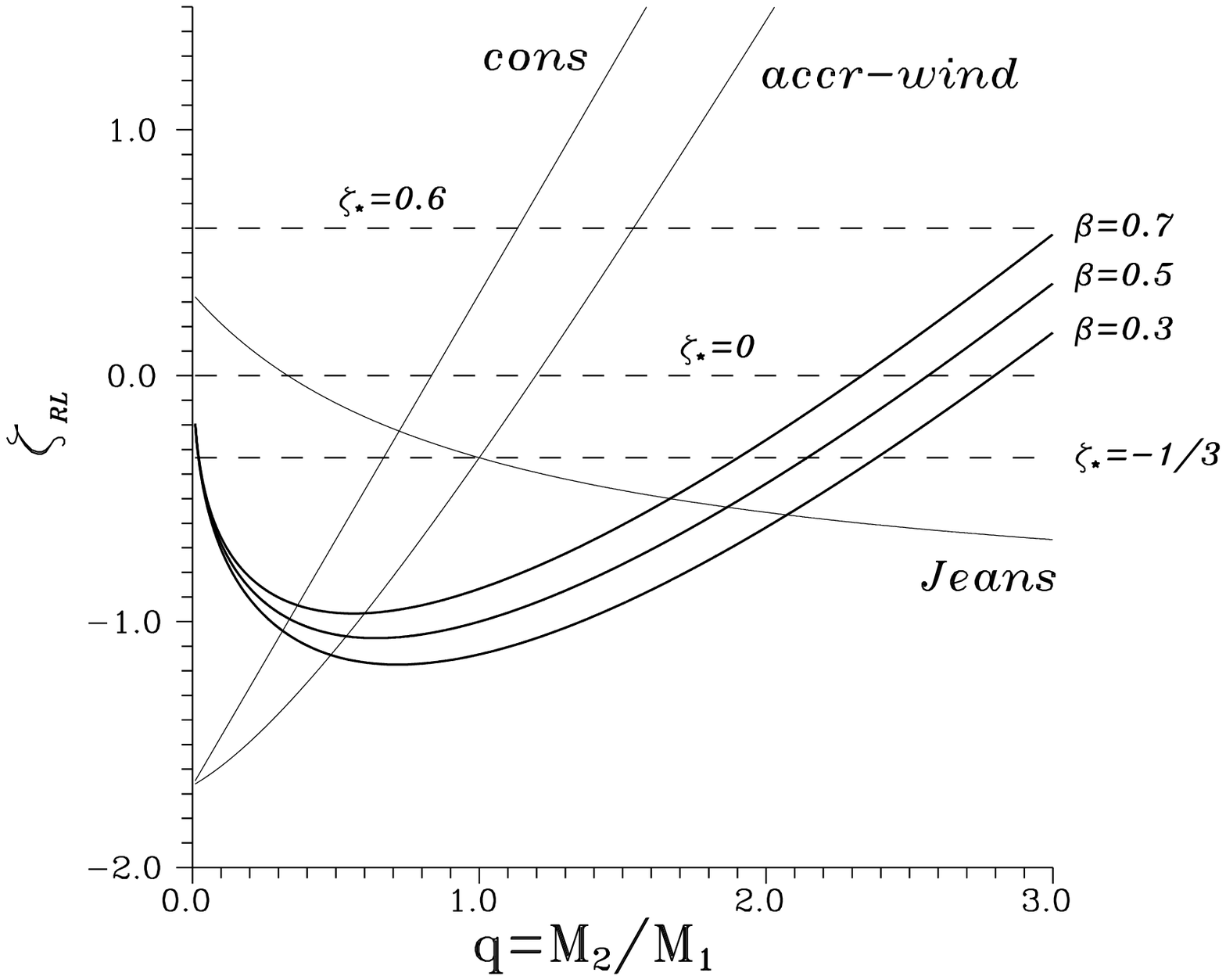,width=16cm}}}
\caption{\small Dependence of logarithmic derivative of
effective radius of Roche lobe w.r.t. the mass of the donor
$\zeta_{RL}=\partial\ln R_{RL}/\partial\ln M_2$ versus mass
ratio $q$ for various values of non-conservativity degree of the
evolution by mass $\beta$ [bold lines; see Eqs~(\ref{drra}
and~(\ref{XXXX})]. Thin lines show dependence $\zeta_{RL}(q)$
for the case of conservative mass exchange [`cons', formula
(\ref{case1})], the case of Jeans mode of mass loss [`Jeans',
formula (\ref{case2})], and the case of stellar wind from
accretor [`accr-wind', formula (\ref{case3})]. Dashed lines show
the values of the logarithmic derivative of the radius of the
donor w.r.t. its mass $\zeta_\star=\partial\ln R_2/\partial\ln
M_2$ for completely  convective and degenerate stars
($\zeta_\star=-\slantfrac{1}{3})$, for thermally equilibrium
stars of MS with $M\sim \Ms$ ($\zeta_\star=0.6)$, and for
subgiants with degenerate low-mass helium cores
($\zeta_\star=0)$.}
\label{ris4}
\end{figure}

We would like to stress also that in the case under
consideration there exists a region of unstable mass exchange
where value of $q$ is very small. The reason of instability
[i.e. the violation of criteria (\ref{dr2r})] is fast increase
of the ratio of the lost angular momentum to the mean specific
momentum of binary system [see Eq.~(\ref{dabpsi}) and
(\ref{XXXX})]. This circumstance can lead to the destruction of
a donor of very low mass.\footnote{Another scenario of
development of mass exchange instability for low $q$ is also
known. It may be caused by low efficiency of interaction between
accretion disk and orbital movement (see, e.g.,
[\ref{rsh83},\ref{hp84}]).} Thus, CVs that begin their evolution
in a stable mode finish it by a  catastrophic disruption of the
donor when  $q$ becomes lower than a certain threshold.

\section{Evolution of CVs with losses of mass and angular
momentum}

To take into account results of 3D gas dynamical simulations of
the mass transfer in CVs and to incorporate proper estimates of
the loss of angular momentum we investigated the  evolution of
CVs on the basis of various assumptions on its conservativity.

Evolution of the donor was calculated using a modified version
of the code for numerical modeling of low-mass stars which was
used in the previous works of the authors (see, e.g.,
[\ref{ft94},\ref{fed97},\ref{fy84}--\ref{tf89}]). The code uses
opacity tables of  Huebner {\it et al.} [\ref{hmma77}] with
addition of data of  Alexander {\it et al.} [\ref{ajr83}] for
low-temperature region. Equation of state was adopted from
Fontaine {\it et al.} [\ref{fgh77}] with modifications suggested
by Denisenkov [\ref{den89}]. Rates of nuclear reaction were
adopted from [\ref{hfcz83},\ref{cfhz85}].

All runs were carried out under assumption that the donor fills
its Roche lobe immediately after arriving on ZAMS. Masses of the
donors were adopted in the range from $0.1\Ms$ to $1.2\Ms$ and
for various values of $q$.  Chemical composition was adopted as
follows: $X=0.70$, $Y=0.28$, $Z=0.02$. For the mixing length
parameter a value of $l/H_p=1.8$ was adopted.  We used the
method of calculation of the donor mass loss rate for known
stellar radius and effective radius of Roche lobe suggested by
Kolb \& Ritter [\ref{kr90}]. Based on the results of gas
dynamical calculations, we assumed that the fraction of matter
accreted by white dwarf is $\beta=0.5$, and that angular
momentum carried away by the matter leaving the system is
defined by expression (\ref{XXXX}). We didn't include into the
model mass and angular momentum loss due to outbursts.

\renewcommand{\thefigure}{5}
\begin{figure}[p]
\centerline{\hbox{\psfig{figure=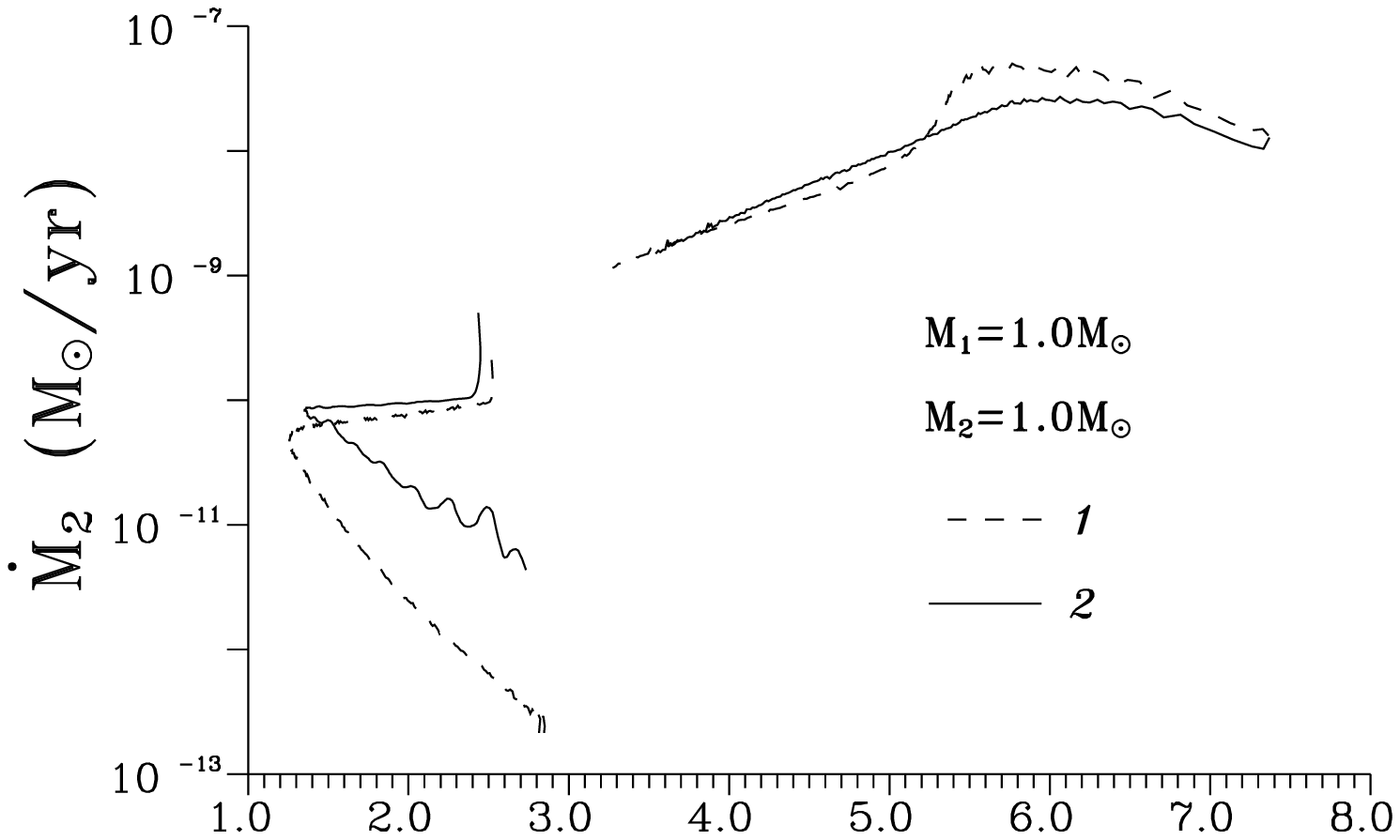,width=15cm}}}
\vspace{5mm}
\centerline{\hbox{\psfig{figure=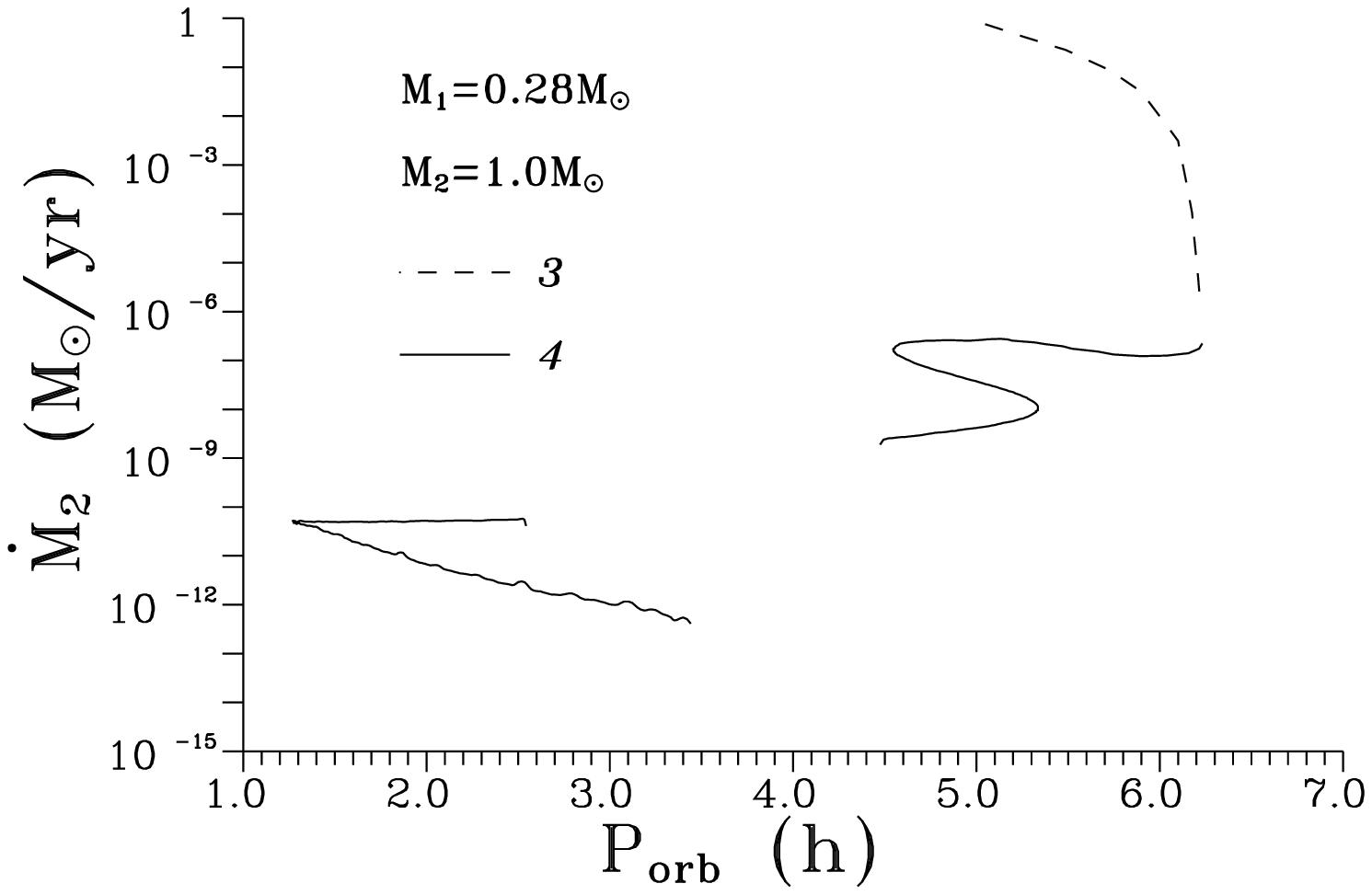,width=15cm}}}
\caption{\small Evolutionary tracks in the ``orbital period --
logarithm of the donor mass loss rate'' plane for binary systems
with initial  masses of components $M_1=1.0\Ms$, $M_2=1.0\Ms$
(top panel) and $M_1=0.28\Ms$, $M_2=1.0\Ms$ (lower panel).
Tracks 1 and 3 (dashed lines) were calculated in the
conservative approximation, and tracks 2 and 4 (solid lines)
were calculated in the  ``non-conservative'' one.}
\label{ris5}
\end{figure}

Let us consider  evolution of CV with parameters corresponding
to the stable mass exchange in conservative model. In
Fig.~\ref{ris5} (``orbital period -- logarithm of donor-star
mass loss rate'' diagram) two evolutionary tracks are shown. One
of the tracks is calculated in the conservative approximation
(track 1), and another one -- in the ``non-conservative''
approximation  (track 2). Initial masses of both donor and
accretor were equal to $1.0\Ms$. It is seen that these two
tracks are almost identical. During the early phase evolution is
driven by angular momentum loss due to the MSW of the donor.
This results in decrease of the orbital semiaxis and orbital
period in the course of evolution. Decrease of the mass of the
donor as well as its radius lead to decrease of $(\partial
A/\partial t)_\MSW$ [see Eq.~(\ref{damsw})], and consequently to
decrease of  mass loss rate by the donor
$\stackrel{\bullet}{M}_2$. For the ``non-conservative'' track,
the value of $\stackrel{\bullet}{M}_2$ is remarkably lower than
for the ``conservative'' one, because in this phase the value of
$\psi$ is negative (see Fig.~\ref{ris2}) and net loss of orbital
angular momentum in the ``non-conservative'' model is lower than
in the conservative one. The situation is inverted in later
phases of evolution as $q$ reaches the values corresponding to
large angular momentum loss.

After the star becomes completely convective,  MSW from the
donor stops.  Mass of the star at this instant is determined by
the deviation from the thermal equilibrium. Single MS stars
become completely convective at $M\sim0.36\Ms$, while
mass-losing components of binaries do at lower masses --
$M\sim0.25\Ms\div0.30\Ms$ (under adopted assumption on chemical
composition and opacity of stellar matter).

In the ``conservative'' track mixing occurs at $M_2=0.265\Ms$
(when the orbital period is equal to
$\Porb=3^{\mbox{h}}\!\!.27$), and in the ``non-conservative''
track -- at $M_2=0.249\Ms$
($\Porb=3^{\mbox{h}}\!\!.54$).\footnote{Since we were interested
in peculiarities of CV evolution that are related to its
possible non conservativity, we formally continued our
calculations beyond accretor mass of  $M_{Ch}=1.4\Ms$ while in
reality in this case the evolution of binary has to be
terminated by a thermonuclear explosion of accretor. The
continuation of calculation for $M_1>M_{Ch}\approx 1.4\Ms$ can
be advocated by the spin-up of accretor due to accretion because
in this case the critical mass can remarkable exceed $M_{Ch}$.}
Thus, values of donor mass and orbital period differ only a
little for the ``conservative'' track and for the
``non-conservative'' one.

The rate of angular momentum loss  decreases abruptly and
contraction of binary systems becomes slower after the donor MSW
is stopped. As a result donor contracts  ceases to lose mass.
Since at the moment of cessation of mass loss the radius of the
donor is larger than the  radius of thermally equilibrium MS
star, the star contracts to the equilibrium radius.  As a result
the ratio of stellar radius to the effective radius of Roche
lobe decreases even more. In this phase the evolution of binary
system is determined by GWR only. During this detached phase the
binary can not be identified as CV, and certain  range of the
orbital periods becomes devoid of stars. This explains the
origin of so called period gap of CVs.

Observed period gap of CVs lies between $2^{\mbox{h}}\!\!.1$ and
$3^{\mbox{h}}\!\!.1$ [\ref{cat97}]. The boundaries of the
``theoretical'' period gap depend on adopted rate of angular
momentum loss due to MSW and on input physics parameters for
evolutionary code that determine theoretical radii of stars. We
used a code that gives somewhat overestimated value for the
lower boundary. Gap boundaries for the ``conservative'' track
are equal to $2^{\mbox{h}}\!\!.5 \div 3^{\mbox{h}}\!\!.3$, and
for the ``non-conservative'' one they are equal to
$2^{\mbox{h}}\!\!.4 \div 3^{\mbox{h}}\!\!.5$. This discrepancy
is related to the  different rate of angular momentum loss just
prior to gap, and with different masses of accretors after
termination  of semi-detached phase of evolution as well.

Note that the period gap is characterized not by total absence
 of CVs but rather by deficiency of CVs. The  presence of CVs in
the period gap can be explained as follows: i) for CVs with
masses of donors equal to $0.25\Ms\div0.4\Ms$ mass exchange
begins immediately in gap; ii) in the case when the donor fills
its Roche lobe not on ZAMS but later, when hydrogen is almost
burn out, then as mass decreases down to $\sim0.3\Ms$ complete
mixing does not occur due to presence a helium core in the star
(see, e.g., [\ref{tfey87},\ref{fe89}]). Though, it must be
admitted  that the number of such systems is very small.

The phase of evolution after period gap is characterized by
lower rate of mass loss by the donor, because evolution now  is
driven by GWR. In this phase the rate of angular momentum loss
from the system is sufficiently lower  compared to the donor MSW
driven phase of evolution.  Another feature of this phase of
evolution is existence of a minimum period of binary.
Short-period cutoff is related to the onset of significant
degeneracy of the material of the donor  and consequently to the
change of mass-radius dependence law [\ref{f71},\ref{p81}]: for
small masses of donors $M_2\sim0.05\Ms$ their  radii increase as
mass decreases and the exponent of the  mass-radius law tends to
$-\slantfrac{1}{3}$.  Minimal period corresponds to the value of
exponent equal to $+\slantfrac{1}{3}$.\footnote{This follows
from approximation (\ref{rrp}) for the effective radius of Roche
lobe $R_{RL}\propto A(M_2/M)^{1/3}$, condition $R_2\approx
R_{RL}$, third Kepler's law $P_{orb}\propto A^{3/2}M^{-1/2}$,
and definition of $\zeta_\star$ (\ref{dr2r}). As a result we
have:
$$
\stackrel{\bullet}{P}/\stackrel{\vphantom{\bullet}}{P}\,=
\slantfrac{3}{2}\stackrel{\bullet}{A}/
\stackrel{\vphantom{\bullet}}{A}\,
-\slantfrac{1}{2}\stackrel{\bullet}{M}/
\stackrel{\vphantom{\bullet}}{M}\,,
$$
$$
\stackrel{\bullet}{R}_2/\stackrel{\vphantom{\bullet}}{R}_2\,=
\slantfrac{1}{3}\stackrel{\bullet}{M}_2/
\stackrel{\vphantom{\bullet}}{M}_2\,
-\slantfrac{1}{3}\stackrel{\bullet}{M}/
\stackrel{\vphantom{\bullet}}{M}\,
+\stackrel{\bullet}{A}/\stackrel{\vphantom{\bullet}}{A}\,,
$$
$$
\stackrel{\bullet}{R}_2/\stackrel{\vphantom{\bullet}}{R}_2\,=
\zeta_\star\stackrel{\bullet}{M}_2/
\stackrel{\vphantom{\bullet}}{M}_2\,,
$$
and, finally
$$
\stackrel{\bullet}{P}/\stackrel{\vphantom{\bullet}}{P}\,\propto
\left(\zeta_\star-\slantfrac{1}{3}\right)
\stackrel{\bullet}{M}_2/\stackrel{\vphantom{\bullet}}{M}_2\,.
$$
This conclusion is independent of the assumption of
conservativity or non-conservativity of mass exchange.} For
``conservative'' track the minimum period is equal to 75 minutes
and for ``non-conservative'' one it is equal to 81 minutes.  The
difference in the values of $P_{min}$ is related to the
difference in the total mass of the system and to the
difference in the deviations of the radii of donor stars from
thermal equilibrium values. The latter deviations are caused by
different rates of mass loss. After passing the minimum the
orbital period begins to increase, and in this phase the rate of
mass loss by the donor quickly decreases.  It is seen from
Fig.~\ref{ris5} that difference between ``conservative''  and
``non-conservative'' tracks becomes larger in this stage. This
can be  naturally explained by increase of the rate of angular
momentum loss with decrease of $q$ in the ``non-conservative''
case.

``Conservative''  and ``non-conservative'' tracks have
appreciably different values of mass transfer rates after
passing the minimum of orbital period. This fact is of
importance since the probability of discovery of  a CV in a
given range of $\log P$ is proportional to

\begin{equation}
p(\log P)~\propto~\frac
{\left(-\stackrel{\bullet}{M}_2\right)^{\displaystyle\gamma}}
{\stackrel{\bullet}{P}/\stackrel{\vphantom{\bullet}}{P}}\,,
\label{ref:prob}
\end{equation}
where $\gamma$ is a positive constant $\simeq1$\ for visual
magnitude limited sample [\ref{kkr98}].

After passing the minimum of the orbital period (i.e. in period
range $2^{\mbox{h}}\div2^{\mbox{h}}\!\!.5 $) the mean value of
the probability of discovery of CV  for the ``non-conservative''
track is 1.2 times larger than for the ``conservative'' one.
Thus, despite a substantial difference in the rate of mass
exchange, the probability of discovery of CV  differs only a
little in two cases under consideration. This is due to faster
crossing of the corresponding period range in the
``non-conservative'' case. Nevertheless, predicted difference in
$\stackrel{\bullet}{M}_2$ for ``conservative'' and
``non-conservative'' cases could be reflected in the distribution
of CVs over types of variability since the character of
variability of a CV depends on the rate of mass exchange.

Figure~\ref{ris5} also shows the evolutionary tracks for stars
that would be dynamically unstable within the conservative
model.  Initial masses of the components were equal to $1.0\Ms$
and $0.28\Ms$. Of course, this combination of masses is rather
extreme but we adopted these parameters for the sake of
illustration.

Track 3 was calculated under the conventional assumptions on the
loss of angular momentum due to GWR and MSW only. This is a
typical case of unstable mass exchange with unlimited growth of
mass exchange rate and fast convergence of the time scale of
mass loss to the dynamical one, i.e. runaway. Strictly speaking
this run should be considered as purely illustrative because in
this case a common envelope has to form  that increases the rate
of angular momentum loss even more (it was not taken into
account in calculations).

Track 4 was calculated within the model in which the  binary
loses matter and angular momentum in accordance with
Eq.~(\ref{XXXX}). It resembles the tracks for conventional case
of stable evolution of CV and differs slightly from track 2 by a
higher rate of mass loss in the initial phase of evolution. The
latter is caused  by some increase of $(\partial A/\partial
t)_\MSW$ at large $q$ [see Eq.~(\ref{damsw})]. It results in
more strong deviations of the donor from thermal equilibrium and
corresponding change of the upper border of period gap: it is
equal to $4^{\mbox{h}}\!\!.4$ for track 4, and
$3^{\mbox{h}}\!\!.5$ for track 2. The minimum of period for
track 4 is equal to 76 minutes, i.e. it is nearly  the same as
for track 1 but slightly less than for track 2.

\section{Boundary of stable mass exchange region}

Boundaries of stable mass exchange region were determined as
follows: several sets of evolutionary tracks were calculated for
systems with predefined masses of donors and  values of $q$ that
were varied with step 0.1. The value of $q$ was considered as
critical if for a given evolutionary track the  mass accretion
 rate didn't exceed Eddington's limit and if for the next
evolutionary track (with  $q$ higher by 0.1) it did. Note that
consequent increasing of $q$ leads to unlimited growth of mass
loss rate (at values of $q$ greater by 0.2 -- 0.3 than
critical). Note also, that results of such procedure are in good
agreement with technique based on comparison of derivatives of
stellar radius and effective Roche lobe radius.

\renewcommand{\thefigure}{6}
\begin{figure}
\centerline{\hbox{\psfig{figure=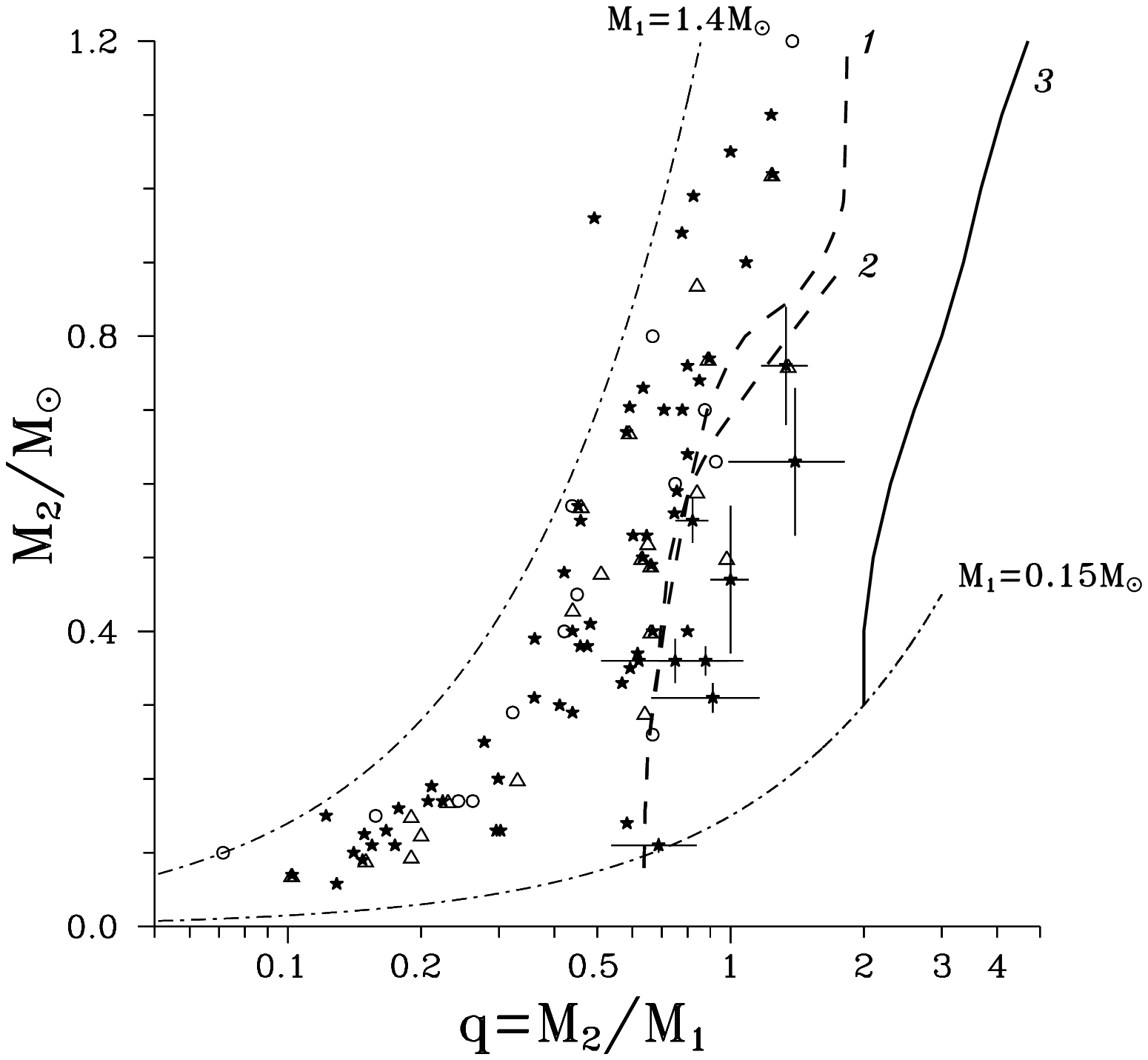,width=17cm}}}
\caption{\small The same as in Figs~\ref{ris1} and~\ref{ris1}a
but with boundary of stable mass exchange region found out in
the present work (bold line 3).}
\label{ris1new}
\end{figure}

The new boundary of the stable mass exchange region is shown in
Fig.~\ref{ris1new} (under assumption of ``non-conservative''
evolution). It is well seen that the new region of stable mass
exchange contains all observed CVs with estimated masses of
components. Thus, within the ``non-conservative'' model the
problem of location of observed systems outside the region of
stable mass exchange does not arise.

\renewcommand{\thefigure}{7}
\begin{figure}[p]
\centerline{\hbox{\psfig{figure=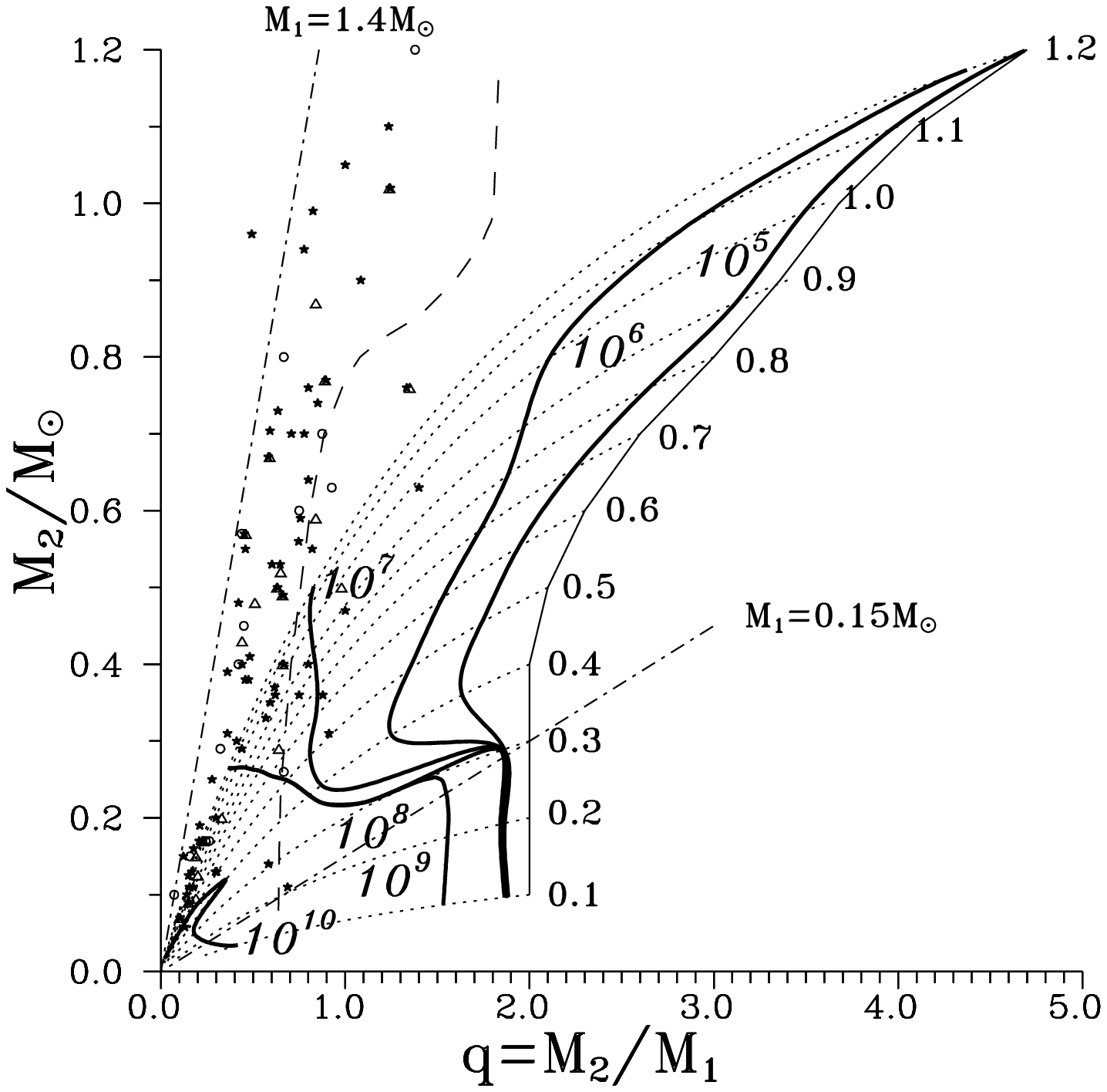,width=16cm}}}
\caption{\small Evolutionary tracks in the ``donor to accretor
mass ratio $q$ -- donor mass $M_2$'' plane for CVs with
different initial donor masses for the ``non-conservative''
model (dotted lines, a number near the beginning of each track
is initial mass of star in $\Ms$). Bold lines are isochrones
with times in years. Thin dashed line -- boundary of the region
of stable mass exchange in the conservative model according to
[\ref{tfy82}], solid line -- the same boundary for
``non-conservative'' model proposed in present work. The
short-dashed lines show the upper ($M_1\leq1.4{M_{\odot}}$) and
lower ($M_1\geq0.15{M_{\odot}}$) cutoffs of white dwarf mass.
Asterisks $\star$ and circles $\circ$ show observed stars from
the catalog of Ritter \& Kolb [\ref{cat97}]. Triangles
$\bigtriangleup$ show observed stars from
[\ref{smith_dhillon}].}
\label{ris7}
\end{figure}

On the other hand, in the $q-M_2$ diagram all observed CVs are
concentrated near $q\aplt 1$, while ``non-conservative'' model
supposes sufficiently higher values of $q$. To investigate this
phenomenon, we have calculated a set of ``non-conservative''
tracks with initial donor masses from $M_2^{init}=0.1\Ms$ up to
$M_2^{init}=1.2\Ms$ with step $0.1\Ms$ and with such values of
$q$ that tracks start in the vicinity of the new boundary of
stable mass exchange region. These tracks are shown in
Fig.~\ref{ris7} ($q-M_2$ diagram). Based on calculations of
these tracks, it is possible to get isochrones that are also
shown in Fig.~\ref{ris7} for the moments of time equal to
$10^5$, $10^6$, $10^7$, $10^8$, $10^9$, $10^{10}$ years (time is
measured from the onset of mass exchange between
components).\footnote{Note, that the isochrones for $10^9$ years
and $10^8$ years for tracks with initial donor mass
$M_2^{init}\ge0.4M_\odot$ merge, since this interval of time
binaries spend in the period gap.} It is this ``extreme'' set of
tracks that was chosen since these systems spent the majority of
time between an old boundary of the stable mass exchange region
and the new one. It is clear, that systems with smaller values
of $q$ (tracks of which  begin to the left from new boundary)
spent there less time.  Thus, CVs with parameters that are close
to the ``extreme'' ones have  most chances to be found out in
this region.

With respect to the location of isochrones in Fig.~\ref{ris7}
all tracks can be subdivided into two distinct groups: i) tracks
for systems with massive donors (from $0.4\Ms$ up to $1.2\Ms$),
for which the evolution prior to the period gap is defined
mainly by the loss of systemic angular momentum due to the donor
MSW; ii) tracks for systems of low-mass donors (from $0.1\Ms$ up
to $0.3\Ms$), for which this stage is absent, because their
donors are initially completely convective.  For the tracks of
the first group a high ($\sim10^{-6}\div10^{-9}$\,\Msyr) rate of
mass loss by the donor prior the period gap is typical, and,
hence, donor mass and mass ratio decrease rapidly. This results
in a fast evolution in the $q-M_2$ diagram. As a result, the
part of the diagram near the new boundary is crossed very
quickly. This explains the absence of observed CVs near new
boundary. For the tracks of the second group (dropping a short
initial phase with large rate of mass exchange) the evolution of
binary system is characterized by low and nearly constant mass
loss rate ($\sim10^{-10}$\,\Msyr) for a long time until period
minimum. Consequently, the mass of the donor and mass ratio of
components change slowly.  Hence, if beginning the evolution
near the  new boundary of stable mass exchange region, the
system should remain there long enough. The absence of CVs in
this region can be explained as follows: masses of dwarfs in
zero-age CVs can not be higher than $\sim0.15\Ms$ that is
minimum mass of helium core for a star leaving MS.

\section{Conclusions}

Results of three-dimensional gas dynamical simulations of the
flow structure in semi-detached binaries show that during mass
exchange a significant fraction of the matter leaves the system.
On the other hand, a number of observed CVs have a combination
of donor mass and mass ratio of components forbidden in the
evolutionary models based on conventional assumptions on the
loss of angular momentum of binaries, since in these cases mass
exchange should be unstable. This contradiction can be resolved
in the framework of the model suggested here.  This model takes
into account losses of mass and angular momentum from the system
according to the results of gas dynamical simulations. It is
shown that approximation for outflowing angular momentum as the
difference of specific angular momentum of the donor matter in
$L_1$ and specific angular momentum of accreted matter (being
multiplied by appropriate mass flows) permits to explain
observations satisfactorily. Note, that the new
``non-conservative'' model practically does not change such
parameters of evolutionary tracks as limits of period gap and
minimal period.

Thus, the main conclusion of our study is: if we use a
``non-conservative'' approximation for CV evolution, we obtain a
new boundary of the region of stable mass exchange in $q-{M_2}$
plane which explains the distribution of observed CVs better
than conventional models.

\section*{ACKNOWLEDGMENTS}

This work was supported by the Russian Foundation for Basic
Research grants 99-02-17619 and 99-02-16037 and Russian
Federation Presidential grant 99-15-96022.

\section*{REFERENCES}

\footnotesize
\begin{enumerate}

\item\label{RJW82} {\it Rappaport, S., Joss, P.C., {\rm\&}
Webbink, R.F.} // Astrophys. J. 1982. V.254. P.616.

\item\label{RVJ83} {\it Rappaport, S., Verbunt, F., {\rm\&}
Joss, P.C.} // Astrophys. J. 1983. V.275. P.713.

\item\label{it84} {\it Iben, I.,Jr., {\rm\&} Tutukov, A.V.} //
Astrophys. J. 1984. V.284. P.719.

\item\label{p84} {\it Patterson, J.} // Astrophys. J. Suppl.
1984. V.54. P.443.

\item\label{w85} {\it Webbink, R.F.} // Interacting Binary Stars
/ Eds W.H.G. Lewin, J. van Paradijs, E.P.J. van den Heuvel.
Cambridge: Cambridge Univ. Press, 1985. P.39

\item\label{kolb93} {\it Kolb, U.} // Astron. Astrophys. 1993.
V.271. P.149

\item\label{ft94} {\it Fedorova, A.V., {\rm\&} Tutukov, A.V.} //
Astron. Zhurn. 1994. V.71. P.431 (Astron. Reports V.38. P.377).

\item\label{rit95} {\it Ritter, H.} //
Evolutionary Processes in
Close Binary Stars / Eds R.A.M.J. Wijers, M.B. Davies, C.A.
Tout. Dordrecht: Kluwer, 1992. P.307.

\item\label{pol96} {\it Politano, M.} // Astrophys. J. 1996.
V.465. P.338.

\item\label{fed97} {\it Fedorova, A.V.} // Binary Stars / Ed.
A.G. Masevich. Moscow: Cosmoinform, 1997. P.179 (in Russian).

\item\label{kmg62} {\it Kraft, R.P., Mathews, J., Greenstein,
J.L.} // Astrophys. J. 1962. V.136. P.312.

\item\label{p67} {\it Paczy\'nski, B.} // Acta Astron. 1967.
V.17. P.267.

\item\label{sch62} {\it Schatzman, E.} // Ann. d'Astroph. 1962.
V.25. P.18.

\item\label{mes68} {\it Mestel, L.} // Mon. Not. R. Astron. Soc.
1968. V.138. P.359.

\item\label{egg76} {\it Eggleton, P.P.} // Structure and
Evolution of Close Binary Stars / Eds P.P. Eggleton, S. Mitton,
J. Whelan. Dordrecht: Reidel, 1976. P.209.

\item\label{vz81} {\it Verbunt, F., {\rm\&} Zwaan, C.} //
Astron. Astrophys. 1981. V.100. P.L7.

\item\label{yu73} {\it Yungelson, L.R.} // Nauchn. Inform. 1973.
V.26. P.71 (in Russian).

\item\label{KIEBOOM81} {\it Kieboom, K.H., {\rm\&} Verbunt, F.}
// Astron. Astrophys. 1981. V.395. P.L11.

\item\label{kk95} {\it King, A., {\rm\&} Kolb, U.} // Astrophys.
J. 1995. V.439. P.330.

\item\label{bbkc98b} {\it Bisikalo, D.V., Boyarchuk, A.A.,
Chechetkin, V.M., Kuznetsov, O.A., {\rm\&} Molteni, D.} // Mon.
Not. R. Astron. Soc. 1998. V.300. P.39.

\item\label{bbkc98c} {\it Bisikalo, D.V., Boyarchuk, A.A.,
Kuznetsov, O.A., {\rm\&} Chechetkin, V.M.} // Astron. Zhurn.
1998. V.75. P.706 (Astron. Reports V.42. P.621; preprint
astro-ph/9806013).

\item\label{eggl83} {\it Eggleton, P.P.} // Astrophys. J. 1983.
V.268. P.368.

\item\label{pacyn71} {\it Paczy\'nski, B.} // Ann. Rev. Astron.
Astrophys. 1971. V.9. P.183.

\item\label{kuz95} {\it Kuznetsov, O.A.} // Astron. Zhurn.
1995. V.72. P.508 (Astron. Reports V.39. P.450)

\item\label{ll62} {\it Landau, L.D., {\rm\&} Lifshitz, E.M.} The
Classical Theory of Fields. 1971, Oxford: Pergamon Press

\item\label{skum72}{\it Skumanich A.} // Astrophys. J. 1972.
V.171. P.565.

\item\label{sr83} {\it Spruit, H.C., {\rm\&} Ritter, H.} //
Astron. Astrophys. 1983, V.124. P.267.

\item\label{hkn96} {\it Hachisu, I., Kato, M., {\rm\&} Nomoto,
K.} // Astrophys. J. Lett. 1996. V.470. P.L97

\item\label{yl98} {\it Yungelson, L., {\rm\&} Livio, M.} //
Astrophys. J. 1998, V.497. P.168.

\item\label{kh94} {\it Kato, M., {\rm\&} Hachisu, I.} //
Astrophys. J. 1994. V.437. P.802.

\item\label{yuetal96} {\it Yungelson, L., Livio, M., Truran, J.,
Tutukov, A., {\rm\&} Fedorova, A.} // Astrophys. J. 1996. V.466.
P.890.

\item\label{lgr91} {\it Livio, M., Govarie, A., {\rm\&} Ritter,
H.} // Astron. Astrophys. 1991. V.246. P.84.

\item\label{skr98} {\it Schenker, K., Kolb, U., {\rm\&} Ritter,
H.} // Mon. Not. R. Astron. Soc. 1998. V.297. P.633.

\item\label{ty71} {\it Tutukov, A.V., {\rm\&} Yungelson, L.R.}
// Nauchn. Inform. 1971. V.20. P.88 (in Russian).

\item\label{tfy82} {\it Tutukov, A.V., Fedorova, A.V., {\rm\&}
Yungelson, L.R.} // Pis'ma Astron. Zhurn. 1982. V.8. P.365 (Sov.
Astron. Lett. V.8. P.198).

\item\label{kool92} {\it de Kool, M.} // Astron. Astrophys.
1992. V.261. P.188.

\item\label{cat97} {\it Ritter, H., {\rm\&} Kolb, U.} // Astron.
Astrophys. Suppl. 1998. V.129. P.83.

\item\label{smith_dhillon} {\it Smith, D.A., {\rm\&} Dhillon,
V.S.} // Mon. Not. R. Astron. Soc. 1998. V.301. P.767.

\item\label{sawada84} {\it Sawada, K., Hachisu, I., {\rm\&}
Matsuda, T.} // Mon. Not. R. Astron. Soc. 1984. V.206. P.673

\item\label{matese83} {\it Matese, J.J., {\rm\&} Whitmire, D.P.}
// Astrophys. J. 1983. V.266. P.776.

\item\label{bbkc98d} {\it Bisikalo, D.V., Boyarchuk, A.A.,
Kuznetsov, O.A., {\rm\&} Chechetkin, V.M.} // Astron. Zhurn.
1999. V.76. P.270 (Astron. Reports V.43. P.229; preprint
astro-ph/9812484).

\item\label{rsh83} {\it Ruderman, M.A., {\rm\&} Shaham, J.} //
Nature. 1983. V.304. P.425.

\item\label{hp84} {\it Hut, P., {\rm\&} Paczy\'nski, B.} //
Astrophys J. 1984. V.284. P.675.

\item\label{fy84} {\it Fedorova, A.V., {\rm\&} Yungelson, L.R.}
// Astrophys. Space Sci. 1984. V.103. P.125.

\item\label{tfey85} {\it Tutukov, A.V., Fedorova, A.V., Ergma,
E.V., {\rm\&} Yungelson, L.R.} // Pis'ma Astron. Zhurn. 1985.
V.11. P.123 (Sov. Astron. Lett. V.11. P.52).

\item\label{tfey87} {\it Tutukov, A.V., Fedorova, A.V., Ergma,
E.V., {\rm\&} Yungelson, L.R.} // Pis'ma Astron. Zhurn. 1987.
V.13. P.780 (Sov. Astron. Lett. V.13. P.328).

\item\label{fe89} {\it Fedorova, A.V., {\rm\&} Ergma, E.V.} //
Astrophys. Space Sci. 1989. V.151. P.125.

\item\label{tf89} {\it Tutukov, A.V., {\rm\&} Fedorova, A.V.} //
Astron. Zhurn. 1989. V.66. P.1172 (Sov. Astron. V.33. P.606).

\item\label{hmma77} {\it Huebner, W.F., Merts, A.L., Magee,
N.H., {\rm\&} Argo, M.F.} // Astrophysical Opacity Library: Los
Alamos Sci. Lab. rep. LA-6760-M. Los Alamos, 1977.

\item\label{ajr83} {\it Alexander, D.R., Johnson, H.R., {\rm\&}
Rypma, R.L.} // Astrophys. J. 1983. V.272. P.773.

\item\label{fgh77} {\it Fontaine, G., Graboske, H.C., {\rm\&}
van Horn, H.M.} // Astrophys. J. Suppl. 1977. V.35. P.293.

\item\label{den89} {\it Denisenkov, P.A.} // Nauchn. Inform.
1989. V.67. P.145 (in Russian).

\item\label{hfcz83} {\it Harris, M.J., Fowler, W.A., Caughlan,
G.R., {\rm\&} Zimmerman, B.A.} // Ann. Rev. Astron. Astrophys.
1983. V.21. P.165.

\item\label{cfhz85} {\it Caughlan, G.R., Fowler, W.A., Harris,
M.J., {\rm\&} Zimmerman, B.A.} // Atom. Data and Nucl. Data
Tabl. 1985. V.32. P.197.

\item\label{kr90} {\it Kolb, U., {\rm\&} Ritter, H.} // Astron.
Astrophys. 1990. V.236. P.385.

\item\label{f71} {\it Faulkner, J.} // Astrophys. J. Lett. 1971
V.170. P.L99.

\item\label{p81} {\it Paczy\'nski, B.} // Acta Astron. 1981.
V.31. P.1.

\item\label{kkr98} {\it Kolb, U., King, A.R., {\rm\&} Ritter,
H.} // Mon. Not. R. Astron. Soc. 1998. V.298. P.L29.

\end{enumerate}

\end{document}